\documentclass{aa}

\usepackage{graphicx}
\usepackage{multirow}
\usepackage{arydshln}
\usepackage{adjustbox}
\usepackage{amsmath}
\usepackage{txfonts}
\usepackage{hyperref}
\usepackage{float}
\hypersetup{
    linkcolor=blue,
    citecolor=blue,
    colorlinks=true,
    filecolor=blue,
    urlcolor=blue}

\begin{document}

   \title{The WEFT project \\
   I. The emergence of turbulence in a cosmic filament}

   \titlerunning{WEFT I}

    \author{Th\'eo Lebeau\inst{1,2} \thanks{Corresponding author: \href{mailto:tlebeau@astro.rug.nl}{tlebeau@astro.rug.nl}}, Saleem Zaroubi\inst{2,1}}

   \authorrunning{Lebeau \& Zaroubi}

   \institute{Kapteyn Astronomical Institute, University of Groningen, Groningen, The Netherlands
        \and
            Astrophysics Research Center of the Open University (ARCO), The Open University of Israel, Israel}

   \date{Received XXX / Accepted XXX}

\abstract
{Much of the ordinary matter in the present-day Universe still escapes direct detection. These missing baryons are thought to reside in a warm-hot intergalactic medium (WHIM) threading the filaments of the cosmic web, whose turbulent and thermal state remains poorly constrained yet underpins any attempt to observe it. Beyond its observational stakes, how this turbulence emerges, as gravitational collapse converts the ordered inflow of the cosmic web into a disordered cascade, is a question of structure formation in its own right.

In this work, we characterise the emergence of turbulence in the diffuse gas of a cosmic filament, following it from its assembly to the present day in the first simulation of the Web Evolution in Filament Targeted zoom simulations (WEFT) project. In this project, the cosmological zoom-in technique is used on a single cosmic filament, which is evolved from $z=63$ to $z=0$ with the moving-mesh code \texttt{AREPO}, reaching a median gas cell size of $\simeq8~\mathrm{kpc}$ within the filament. We trace the assembly of the filament and the thermodynamic state of its diffuse gas, the generation of vorticity at its accretion shocks through the kinetic helicity, enstrophy, and baroclinic term, and the growth of turbulent motions over cosmic time. We then quantify the intermittency of the cascade through high-order velocity structure functions and their relative scaling exponents from extended self-similarity. We show that the filament assembles by the hierarchical merging of several proto-filaments rather than by laminar accretion. We find that vorticity is seeded baroclinically where the rotating strands meet their accretion shocks obliquely. We show that the flow evolves from a supersonic, shock-dominated state, statistically close to the bifractal Burgers limit, into a developed, mildly supersonic, and intermittent cascade whose longitudinal exponents lie closest to the sheet-like She--L\'ev\^eque model at $z=0$. Turbulence is thus an intrinsic and quantifiable property of the diffuse gas of filaments, emerging through baroclinic vorticity generation and maturing into a developed, intermittent cascade. This pathfinder run provides the turbulent input on which forecasts of the observability of the WHIM can be based.}


\maketitle

\section{Introduction}
\label{sec:intro}

Matter in the Universe is arranged in a cosmic web of sheets, filaments, and
nodes \citep{bond1996filaments}. Filaments are the dominant component of this
web, holding roughly half of the mass at the present
day \citep{cautun2014evolution}, and their gas is predominantly warm-hot, at
temperatures of $10^5$--$10^7$~K, the warm-hot intergalactic medium
(WHIM). Simulations predict that 30 to 40 per cent of all baryons reside in this phase globally \citep{1999ApJ...514....1Cen,2001ApJ...552..473Dave}, making it the
leading reservoir for the ordinary matter that has long escaped direct
detection, the missing-baryon
problem \citep{White1993-baryon_fraction_cosmology,2004ApJ...616..643Fukugita,2012ApJ...759...23Shull}.

Observational access to this gas has advanced steadily, outward from the
densest, hottest regions of the cosmic web. As X-ray and Sunyaev--Zel'dovich
techniques matured, they reached progressively cooler and more diffuse gas at
ever larger distances from cluster cores, from the outskirts of galaxy
clusters \citep{ettori2013mass,2013A&A...551A..22Eckert} to the bridges
connecting close cluster pairs \citep{pereyra2020detection,migkas2025detection}
and, through the stacking of many systems, to the filaments
themselves \citep{degraaff2019probing,tanimura2022xray}. The diffuse web is now
being probed in the radio, where stacking experiments have uncovered
synchrotron emission from the filamentary gas \citep{Vernstrom2021,Vernstrom2023}
and a radio ridge has been resolved between a close cluster
pair \citep{Govoni2019}, tracing the relativistic particles accelerated by the
shocks and turbulence that pervade it \citep{BrunettiVazza2020}. More direct
probes are developing in parallel, from redshifted H\,{\sc i} 21~cm
emission \citep{2017MNRAS.468..857Kooistra,2019MNRAS.490.1415Kooistra}, with
intergalactic H\,{\sc i} now detected in nearby
filaments \citep{Arabsalmani2025}, to the dispersion of fast radio bursts used
to weigh the missing baryons \citep{2020Natur.581..391Macquart,2025ApJ...995..183Mo}.

Line-intensity mapping complements these efforts, measuring the cumulative
emission of the unresolved gas rather than individual sources and returning a
statistical, tomographic view of the diffuse
web \citep{Kovetz2017LIM,Tramonte2019,BernalKovetz2022}. Because each spectral
line arises from gas in a particular thermal and ionisation state, different
lines trace the distinct phases that thread filaments, from the H\,{\sc i}
21~cm line and the Lyman-$\alpha$ line to the O\,{\sc vii} and O\,{\sc viii}
X-ray lines that probe the hot, collisionally ionised WHIM, so far seen in
absorption \citep{Nicastro2018}. The Square Kilometre Array (SKA) will extend several
of these efforts \citep{Pan2026SKA,Cuciti2026SKA}. Whether resolved, stacked, or
cumulative, these probes remain largely statistical or indirect, and their
interpretation depends sensitively on the still-uncertain turbulent and
thermal state of the gas.

Resolving that small-scale state has, however, remained out of reach. In large
cosmological volumes filaments are abundant, but their gas is sampled too
coarsely for its internal dynamics to be followed \citep{gheller2019}. Filaments moreover span a heterogeneous population, from short dense bridges to long
tenuous strands \citep{galarraga2020populations}. High-resolution zoom-in
simulations have instead resolved filaments only incidentally, in work
intended to study cold streams and protocluster bridges through which the
high-redshift cosmic web feeds galaxy formation at Cosmic
Noon \citep{2019MNRAS.484.1100Mandelker,2021ApJ...923..115Mandelker,2024MNRAS.52711256L_Lu},
or as structures in the outskirts of galaxy-cluster
zooms \citep{2018MNRAS.480.2898Cui,rost2021the300,sorce2021hydrodynamical,2024A&A...686A.157NelsonTNGCluster}.
The closest of these traced velocity fields and turbulence from filaments to
clusters \citep{2025A&A...704A..14Lebeau}, but at a resolution insufficient to
characterise the cascade within the diffuse gas itself. In none of these cases
is the diffuse filament gas followed as the primary object, nor its turbulence
characterised, to the present day.

This gap matters beyond the baryon census. How gravitational collapse converts
an ordered inflow into a developed turbulent cascade is a question about
structure formation in its own right, and the diffuse filament gas is where it
can be followed in its least-processed form, away from the feedback that
complicates denser environments. The same diffuse gas is also the reservoir
from which the galaxies embedded in the filament accrete, and it pre-processes
those streaming along it towards clusters, through starvation and ram-pressure
effects that begin well outside the cluster
environment \citep{2018MNRAS.474..547Kraljic,gunn1972infall}. Both the inward
flow that feeds these galaxies and the stripping that depletes them depend on
the dynamical and thermal state of the gas, making it a key ingredient of
galaxy evolution in the cosmic web.

Here we present the first simulation of the Web Evolution in Filament Targeted
zoom simulations (WEFT) project, in which the cosmological zoom-in technique is
turned on a cosmic filament rather than on a halo. Using the moving-mesh code
\texttt{AREPO} \citep{2010MNRAS.401..791SpringelAREPO,2020ApJS..248...32Weinberger}, we
follow one such filament from $z=63$ to $z=0$, reaching a median gas cell size
of $\simeq8~\mathrm{kpc}$ within it, from unconstrained initial
conditions \citep{hahn2011multi} and including radiative cooling, a photoionising
ultraviolet background, and star formation with stellar feedback (Sect.~\ref{sec:sim}).
To our knowledge, this pathfinder run is the first cosmological simulation to
resolve the internal gas dynamics of a filament as its primary target down to $z=0$.

This paper is organised as follows. We describe the WEFT pathfinder simulation,
the selection of the diffuse filament gas and the diagnostics used throughout in
Sect.~\ref{sec:methods}. We then present our results, from the assembly of the
filament (Sect.~\ref{sec:assembly}) and the generation of vorticity at its
accretion shocks (Sect.~\ref{sec:vort}) to the growth of turbulence over cosmic
time (Sect.~\ref{sec:growth}) and the intermittency of the developed cascade
(Sect.~\ref{sec:cascade}). We discuss these results in
Sect.~\ref{sec:discussion} and conclude in Sect.~\ref{sec:conclusion}.

\section{Methods}
\label{sec:methods}

\subsection{The WEFT pathfinder simulation}
\label{sec:sim}

The initial conditions of this simulation are generated with the multi-scale code \texttt{MUSIC} \citep{hahn2011multi}, adopting the \textit{Planck} 2018 cosmology \citep{2020A&A...641A...6Planck} ($\Omega_\mathrm{m}=0.310$, $\Omega_\mathrm{b}=0.049$, $\Omega_\Lambda=0.689$, $h=0.677$, $\sigma_8=0.810$, $n_\mathrm{s}=0.967$). We start from a dark-matter-only parent box of $50~\mathrm{Mpc}/h$ on a side, sampled on a $128^3$ root grid, large enough to develop a representative filamentary network while keeping the subsequent zoom affordable. From a first dark-matter-only low-resolution realisation evolved to $z=0$, we select a single, well-defined filament threading two main haloes. Unlike constrained realisations of the local Universe, we impose no external constraint on the density field. This simulation thus follows a generic filament, and the absolute levels we report are subject to cosmic variance (Sect.~\ref{sec:limitations}).

\begin{figure}
     \centering
     \includegraphics[trim=0 60 1100 0,clip, width=0.9\linewidth]{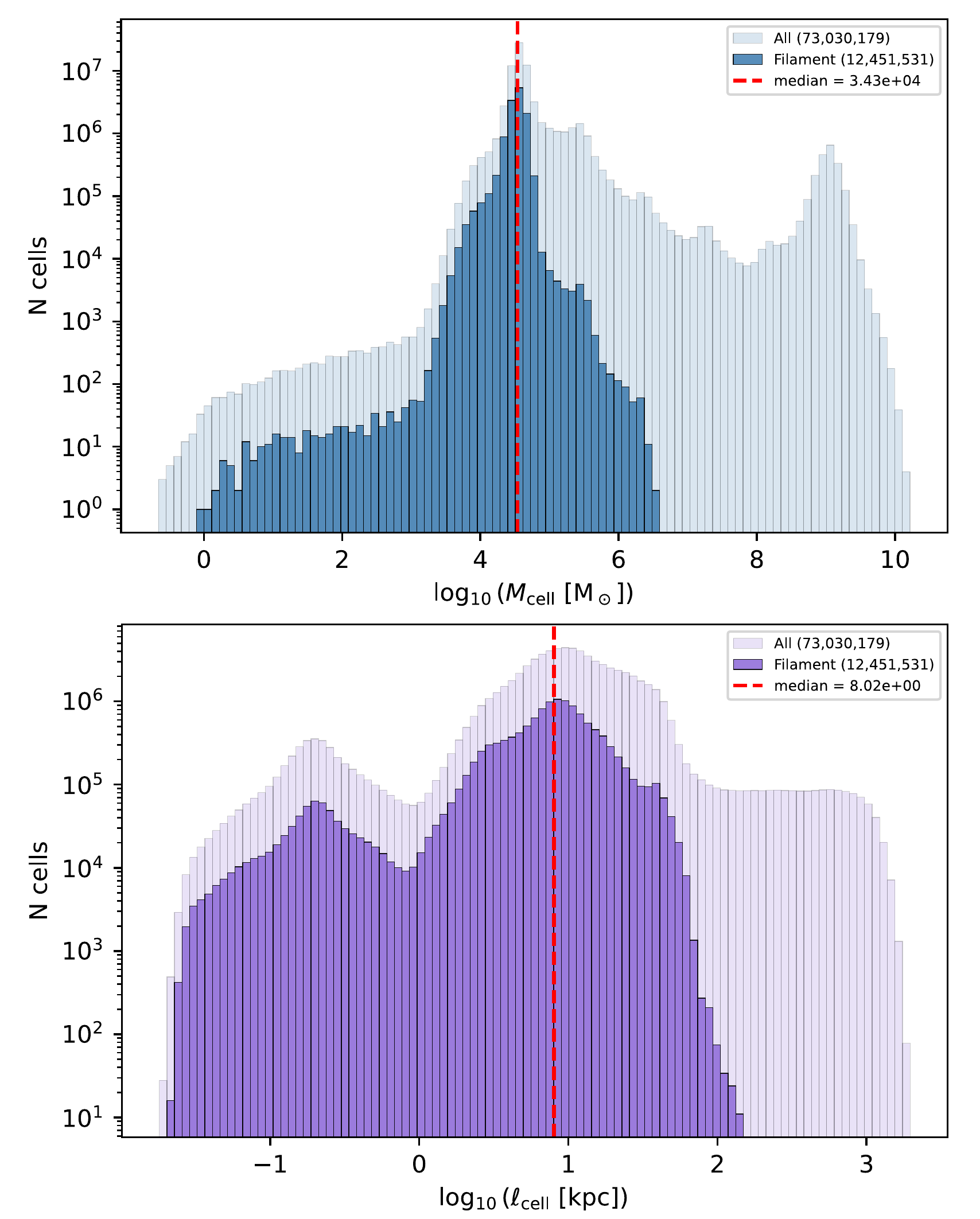}
\caption{Gas cell mass (top) and effective size (bottom) distributions at $z=0$,
for the diffuse cells of the filament region (dark), defined as a $4\times2\times2~\mathrm{cMpc}/h$ region centred on the spine of the filament, and all cells of the
$50~\mathrm{Mpc}/h$ box (light grey).}
\label{fig:hists}
\end{figure}

The Lagrangian volume enclosing the filament and its immediate surroundings at $z=0$ is traced back to the initial conditions at $z=63$, and this region is refined to an effective resolution of $4096^3$. In the zoom region, dark matter particles have a mass $m_{\rm DM}\simeq1.3\times10^{5}\,h^{-1}\,M_\odot$ and use a comoving gravitational softening of $0.25\,h^{-1}\,\mathrm{kpc}$, about $1/49$ of the mean high-resolution interparticle spacing, capped to a maximum physical value of $0.125\,h^{-1}\,\mathrm{kpc}$ below $z=1$. Gas is introduced throughout the high-resolution region with a target cell mass of $2.47\times10^4\,h^{-1}\,M_\odot$ ($3.65\times10^4\,M_\odot$), and uses an adaptive softening set to $2.5$ times the local cell radius, with a comoving floor of $6.25\times10^{-3}\,h^{-1}\,\mathrm{kpc}$. The moving mesh continually refines and de-refines to keep the gas cells near this target mass (top panel of Fig.~\ref{fig:hists}), so that the diffuse filament gas, being overdense, is resolved more finely than the cosmic mean. The moving mesh thus reaches a median gas cell size of $\simeq8~\mathrm{kpc}$ in the filament (see lower panel of Fig.~\ref{fig:hists}), the kiloparsec-scale resolution quoted throughout.

Following standard practice for zoom simulations, the high-resolution region is enlarged and reshaped iteratively so that no low-resolution particle enters the filament over the whole history \citep{1993ApJ...412..455KatzWhite,2014MNRAS.437.1894Onorbe}, while successive shells of coarser particles pad it out to the box scale to represent the large-scale tidal field. Low-resolution dark-matter particles do not penetrate the filament. Within the $4\times2\times2~\mathrm{cMpc}/h$ region analysed in this work, they make up only $0.16\%$ of the particles by number and $1.50\%$ by mass (against $2.67\%$ and $62.27\%$ across the broader zoom region), so the analysed gas is essentially free of contamination.

We evolve the initial conditions from $z=63$ to $z=0$ with the moving-mesh code \texttt{AREPO} \citep{2010MNRAS.401..791SpringelAREPO,2020ApJS..248...32Weinberger}, which solves the equations of hydrodynamics with a finite-volume Godunov scheme on an unstructured Voronoi mesh that moves with the flow. This quasi-Lagrangian discretisation follows the gas without a preferred frame, automatically concentrating resolution in dense regions while remaining accurate across the strong shocks that bound the accreting filament, properties well suited to resolving the internal dynamics of the diffuse gas. The calculation uses the development version of the code.

This pathfinder run includes the radiative physics relevant to the warm-hot intergalactic medium. These comprise primordial radiative cooling, a spatially uniform, redshift-dependent photoionising ultraviolet background \citep{2009ApJ...703.1416FaucherGiguere}, corrected for the self-shielding of dense gas \citep{2013MNRAS.430.2427Rahmati}, and star formation with its associated thermal feedback, treated through the two-phase effective equation-of-state model of \citet{2003MNRAS.339..289SpringelHernquist}. Star-forming gas therefore lies on an effective pressure floor rather than following a fully resolved multiphase interstellar medium. We do not employ the more elaborate galaxy-formation model of the IllustrisTNG suite \citep{2018MNRAS.473.4077Pillepich}, since our target is the diffuse intergalactic gas rather than the internal structure of galaxies. The calculation is purely hydrodynamic and does not include magnetic fields, a limitation we return to in Sect.~\ref{sec:limitations}. We store $118$ snapshots, with the output cadence densified towards low redshift ($\Delta a=0.005$ for $a>0.755$) to resolve the recent evolution of the cascade. The production run reaches $z=0$ in $\simeq4.8\times10^5$~CPU-hours. 

Finally, to test that our results are not resolution artefacts, we repeat the intermittency and vorticity diagnostics on a lower-resolution version of the run, at an effective resolution of $2048^3$ cells, evolved from the same initial conditions with identical physics. The comparison, presented in Appendix~\ref{app:conv}, confirms that the resolution of the primary run is sufficient to resolve the processes studied here.

\subsection{Selecting the diffuse filament gas}
\label{sec:selection}

All of the analysis that follows is carried out within a fixed comoving box of $4\times2\times2~\mathrm{cMpc}/h$ enclosing the filament, which defines the region over which we characterise the diffuse gas. Because the filament is not stationary in the simulation volume, its centre is identified at each epoch from the projected gas density in the $y$--$z$ plane and interpolated smoothly between snapshots, so that the box tracks the same structure throughout its assembly.

A meaningful measurement of the turbulent cascade requires isolating the diffuse intergalactic gas from the collapsed objects embedded in the filament. Inside and around haloes, the velocity increments are set by the deep potential wells of the galaxies and groups, through coherent rotation and bulk infall, which inject large, spatially localised increments that mimic the signature of intermittency. We therefore excise every gas cell lying within $2\,R_{200}$ of a halo more massive than $10^{10}\,M_\odot$, identified with the friends-of-friends and \textsc{Subfind} algorithms \citep{1985ApJ...292..371Davis,2001MNRAS.328..726Springel} and free of low-resolution particle contamination, and we additionally discard the star-forming cells, which lie on the effective equation of state \citep{2003MNRAS.339..289SpringelHernquist} and whose temperature and velocity are sub-grid quantities rather than resolved turbulence. We do not impose a temperature cut, so the sample is the full diffuse gas threading the filament, spanning both the cold photoionised and the warm-hot phases (see the density--temperature phase-space diagram in Fig.~\ref{fig:phase}), as expected for a tenuous filament \citep{galarraga2021properties}. The halo excision and the star-formation cut already remove the collapsed and sub-grid gas whose increments would otherwise mimic intermittency. We characterise this diffuse gas as a whole, disentangling
the turbulent state of its cold and warm-hot phases, and the role of cold
accretion flows, is left to future work in the WEFT programme.

\begin{figure}
     \centering
     \includegraphics[trim=0 0 1280 0, clip, width=0.9\linewidth]{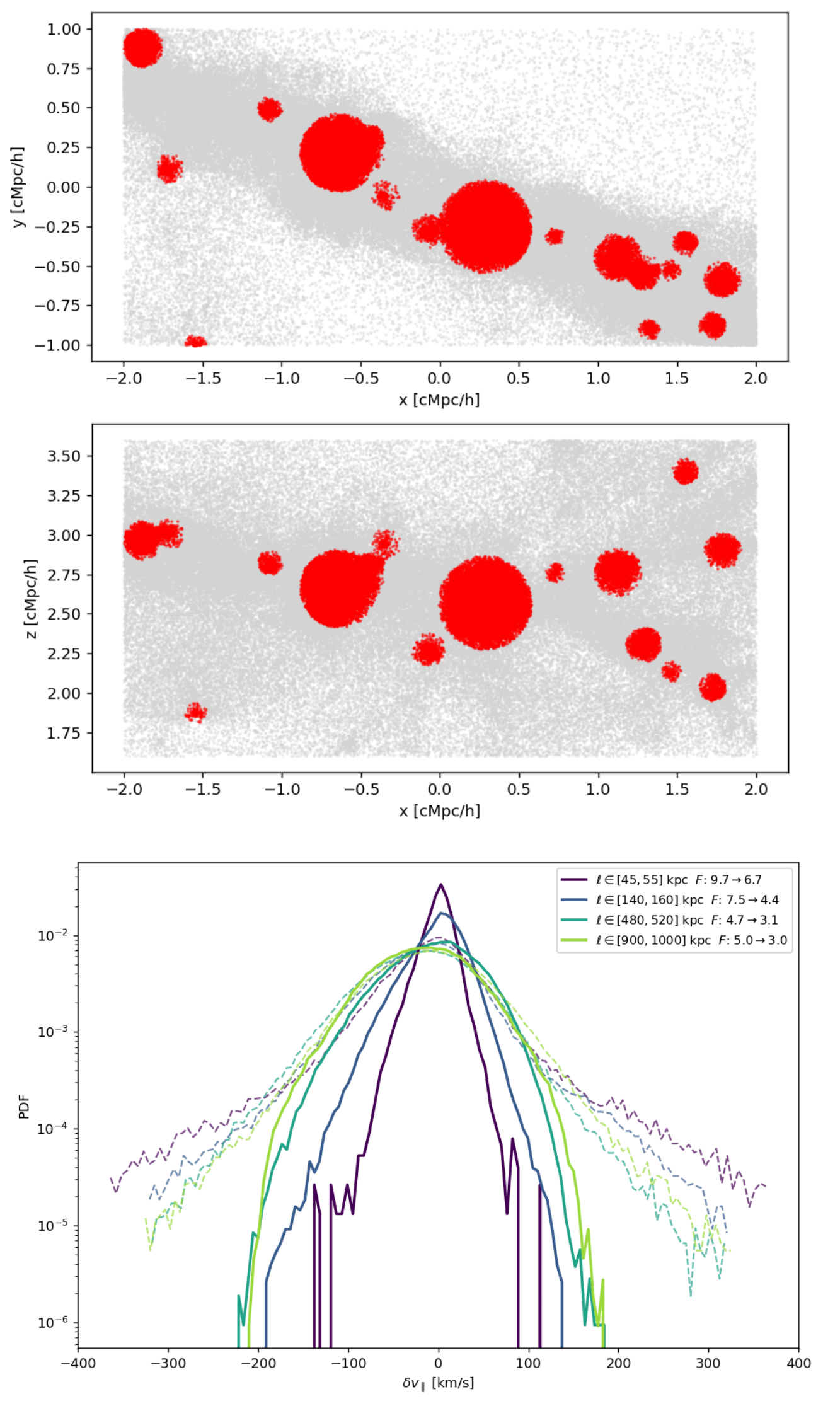}
     \caption{Top: the retained diffuse gas
(grey) and the cells removed by the halo mask (red) in the $x$--$y$ and $x$--$z$
planes. Bottom: probability density functions of the longitudinal increment
$\delta v_\parallel$ at four separations, before (dashed) and after (solid)
masking, with the flatness annotated for each.}
\label{fig:mask}
\end{figure}

This selection removes $\sim37\%$ of the cells in the box, tracing the massive haloes and embedded galaxies while preserving the diffuse gas that threads between them (top panel of Fig.~\ref{fig:mask}). Its effect is clearest in the statistics of the velocity increments (bottom panel of Fig.~\ref{fig:mask}). Before the cut the increment distributions carry heavy, non-Gaussian wings at all separations, with a flatness reaching $F\sim10$ at the smallest scales, while afterwards the wings are strongly suppressed and the distributions relax towards a Gaussian shape ($F\to3$) at large separations, leaving the residual small-scale non-Gaussianity that is the genuine intermittency of the diffuse cascade.

\subsection{Rotation and vorticity diagnostics}
\label{sec:diag}

We characterise the dynamical and thermal state of the diffuse gas through a set of per-cell quantities. First, the temperature follows from the internal energy and electron abundance stored by \texttt{AREPO}, as 

\begin{equation}
    T=(\gamma-1)\,u\,\mu\,m_\mathrm{p}/k_\mathrm{B}
\end{equation}

\noindent with $\gamma=5/3$ the adiabatic index, $u$ the internal energy, $\mu$ the mean molecular weight fixed by the electron fraction, $m_\mathrm{p}$ the proton mass and $k_\mathrm{B}$ the Boltzmann constant. The sound speed is then defined as

\begin{equation}
    c_s=\sqrt{\gamma k_\mathrm{B} T/\mu m_\mathrm{p}} \, .
\end{equation}

The compressive state of the flow is thus measured by the cell Mach number 

\begin{equation}
    \mathcal{M}=|\boldsymbol{v}|/c_s
\end{equation}

\noindent and by the three-dimensional velocity divergence

\begin{equation}
    \nabla\cdot \boldsymbol{v}= \frac{\partial{v_x}}{\partial{x}}+\frac{\partial{v_y}}{\partial{y}}+\frac{\partial{v_z}}{\partial{z}} \, , 
\end{equation}

\noindent whose strongly negative regions mark the converging surfaces of the accretion shocks.

Rotational motions are quantified by the vorticity 

\begin{equation}
\boldsymbol{\omega} = \nabla \times \boldsymbol{v} =
\begin{pmatrix}
\dfrac{\partial v_z}{\partial y} - \dfrac{\partial v_y}{\partial z} \\[2.2ex]
\dfrac{\partial v_x}{\partial z} - \dfrac{\partial v_z}{\partial x} \\[2.2ex]
\dfrac{\partial v_y}{\partial x} - \dfrac{\partial v_x}{\partial y}
\end{pmatrix} ,
\label{eq:vorticity}
\end{equation}

\noindent and the associated enstrophy 

\begin{equation}
    \mathcal{E}=\tfrac12|\boldsymbol{\omega}|^2 \, ,
\end{equation}

\noindent a positive-definite measure of the local rotational energy that traces the developed cascade. The large-scale, coherent rotation of the assembling strands is captured by the kinetic helicity 

\begin{equation}
    \mathcal{H}=\boldsymbol{v}\cdot\boldsymbol{\omega} \, ,
\end{equation}

\noindent which changes sign across the spine and serves as a tracer of that ordered rotation rather than as its cause \citep{2021NatAs...5..839Wang}. Finally, the generation of vorticity is diagnosed through the baroclinic term 

\begin{equation}
    \mathcal{B}=|\nabla\rho\times\nabla P|/\rho^2 \, ,
\end{equation}

\noindent the magnitude of the baroclinic vector that sources vorticity wherever the density and pressure gradients are misaligned, as at an oblique shock \citep{2008Sci...320..909Ryu,2017MNRAS.464..210Vazza,2017MNRAS.464.4448Wittor}. The baroclinic term and the enstrophy share the same dimension of an inverse time squared and are both expressed in $\mathrm{km^2\,s^{-2}\,kpc^{-2}}$, so that they can be compared directly.

The vorticity and divergence follow from the velocity Jacobian, which, together with the density and pressure gradients, is estimated for each cell by a least-squares fit over its $32$ nearest neighbours. Because a derivative taken next to an excised halo or a box edge would be biased by its truncated neighbourhood, these gradients are computed in two stages. We first build a single neighbour tree over a computation region that pads the $4\times2\times2~\mathrm{cMpc}/h$ box by a margin of $800~\mathrm{kpc}/h$ and retains all cells, haloes included, so that every cell in the box keeps its true physical neighbours. Only then do we apply the halo excision and the star-formation cut and average over the selected cells. Because this excision still leaves a few rare, high-vorticity cells that would dominate an arithmetic mean of the rotational diagnostics, we characterise the diffuse gas through the median of the enstrophy, baroclinic term and cell Mach number, which follows the bulk of the volume rather than these outliers. The turbulent Mach number is instead built from the three-dimensional velocity dispersion of the selected gas, 

\begin{equation}
    \sigma_v=\langle|\boldsymbol{v}-\langle \boldsymbol{v}\rangle|^2\rangle^{1/2} \, ,
\end{equation}

\noindent with $\mathcal{M}_\mathrm{turb}=\sigma_v/\langle c_s\rangle$, which isolates the turbulent motions from the coherent large-scale infall still carried by the cell-by-cell Mach number at early times. Because $\sigma_v$ is the full three-dimensional dispersion, the turbulent pressure $P_\mathrm{turb}=\rho\sigma_v^2/3$ compares with the thermal pressure $P_\mathrm{th}=\rho c_s^2/\gamma$ as $P_\mathrm{turb}/P_\mathrm{th}=\gamma\mathcal{M}_\mathrm{turb}^2/3$, which reaches $\simeq2.0$ at $z=0$. The same per-cell quantities are evaluated in every snapshot, on the identical selection, to follow their evolution across cosmic time.

\subsection{Velocity structure functions and intermittency}
\label{sec:vsf}

We quantify the turbulent cascade through the velocity structure functions of the diffuse gas defined as 

\begin{equation}
    S_p(\ell)=\langle|\delta v_\parallel(\ell)|^p\rangle
\end{equation}

\noindent where $\delta v_\parallel=(\boldsymbol{v}_2-\boldsymbol{v}_1)\cdot\hat{\boldsymbol{\ell}}$ is the longitudinal velocity increment between two cells separated by $\boldsymbol{\ell}$, projected onto the separation unit vector, and the average runs over all pairs at a given separation \citep{1941DoSSR..30..301Kolmogorov,1995frischTurbulence}. We compute the orders $p=1$ to $8$. The transverse increments, used to probe the isotropy of the cascade (see Appendix~\ref{app:iso}), are defined analogously in the plane perpendicular to $\boldsymbol{\ell}$. For fully developed, incompressible turbulence the four-fifths law fixes the third-order longitudinal exponent to $\zeta_3=1$ exactly. Although the law is stated for the signed increment $\langle\delta v_\parallel^3\rangle$, the absolute structure functions $S_p=\langle|\delta v_\parallel|^p\rangle$ used here share the same inertial-range scaling, so that $\zeta_3=1$ anchors the relative scaling used below. We nonetheless measure $\zeta_3$ independently, from the slope of $\log S_3$ against $\log\ell$ over the fit range rather than through extended self-similarity, and recover $\zeta_3\simeq0.9$--$1.0$ as the fit window is varied over plausible bounds. The relative exponents quoted below are therefore not anchored to unity by construction.

\begin{figure*}[h]
     \centering
     \includegraphics[width=0.7\linewidth]{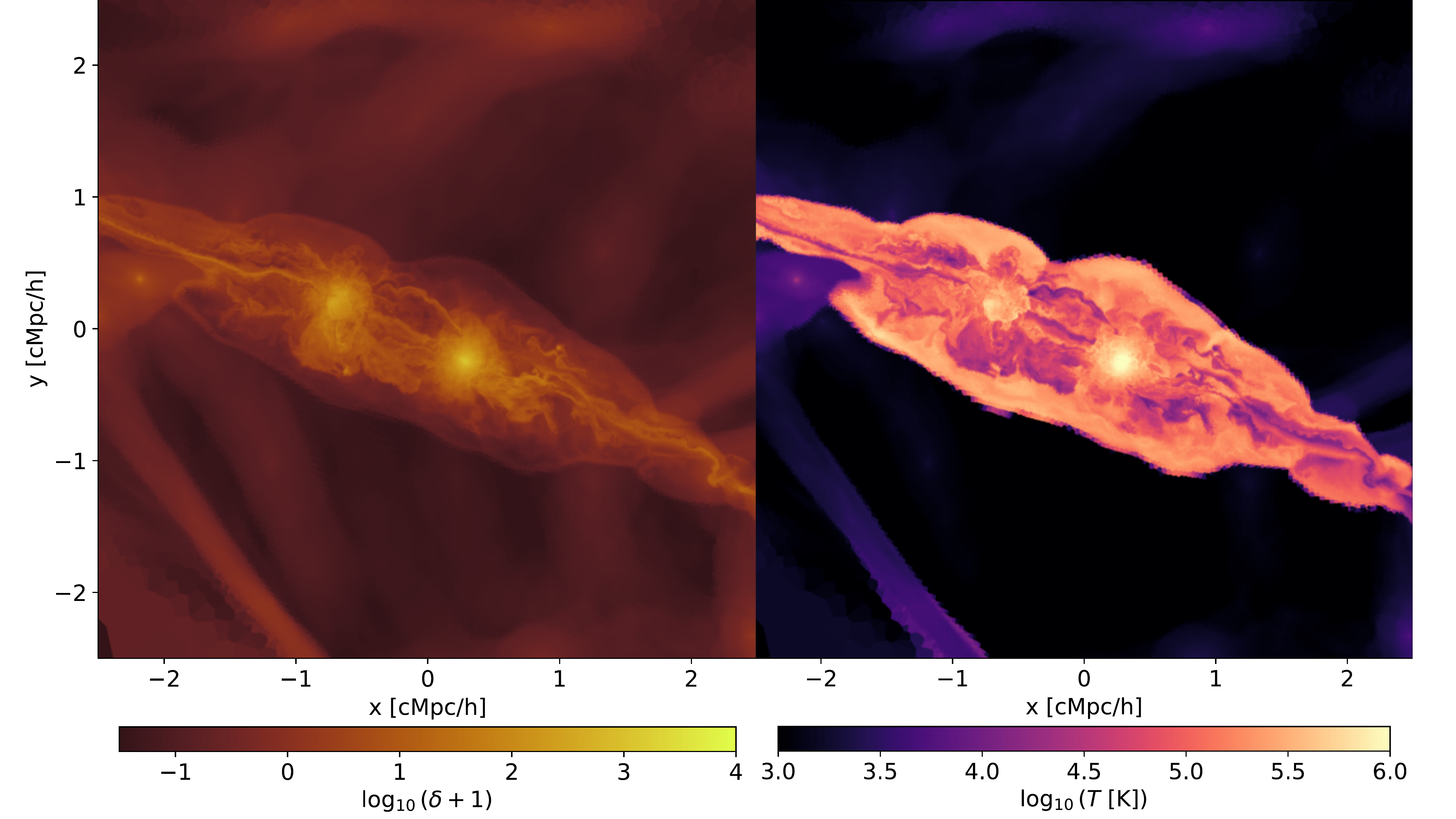}
\caption{Overdensity (left) and gas temperature (right) at $z=0$ in a comoving region of $5\times5~\mathrm{cMpc}/h$ centred on the filament projected along the $z$ axis.}
\label{fig:z=0}
\end{figure*}

A distinctive feature of our measurement is that the structure functions are evaluated directly on the Voronoi cells, using their positions and velocities, without first interpolating the velocity field onto a regular grid. Interpolating a moving-mesh field onto a grid finer than the local cell size produces many pairs of grid points that fall inside the same Voronoi cell and therefore share an identical velocity. These spurious zero-increment pairs artificially depress the high-order increments and flatten the measured exponents, saturating $\zeta_p$ towards unity for $p\gtrsim3$. Working on the native cells removes this bias entirely. To make the pair counting tractable over the $\sim8\times10^6$ cells of the filament region, we accumulate pairs in logarithmically spaced separation bins with a hybrid strategy. At small separations, an exact neighbour search (a chunked \texttt{cKDTree} ball query, binned on the fly) enumerates all pairs up to a transition radius set by the available memory, while at large separations pairs are drawn by distance-filtered random sampling. 

Each bin is filled to $\sim5\times10^5$ pairs. This robustly constrains the exponents up to sixth order, on which our quantitative statements rest, while the seventh and eighth orders, shown for completeness, probe the deepest tails and carry correspondingly larger statistical uncertainty. Direct fits of $\log S_p$ against $\log\ell$, from which we obtain the absolute third-order exponent $\zeta_3$, are performed over the range $[80,700]~\mathrm{ckpc}/h$. Its lower bound is about fifteen times the median comoving cell size in the filament ($\ell_\mathrm{cell}\simeq5.4~\mathrm{ckpc}/h$, i.e. $\simeq8~\mathrm{kpc}$ in physical units at $z=0$), which keeps the fit clear of numerical dissipation, while its upper bound is the injection scale, beyond which the increments are dominated by the large-scale accretion flows and the finite transverse extent of the filament rather than by the cascade. This bounded window is required only for these direct, absolute-scale fits. The relative exponents $\zeta_p/\zeta_3$ introduced below are instead measured over the full sampled range through extended self-similarity. The structure functions follow clean power laws over the fit range (see Fig.~\ref{fig:vsf}).

\begin{figure*}
     \centering
     \includegraphics[trim= 0 600 0 0, clip, width=\linewidth]{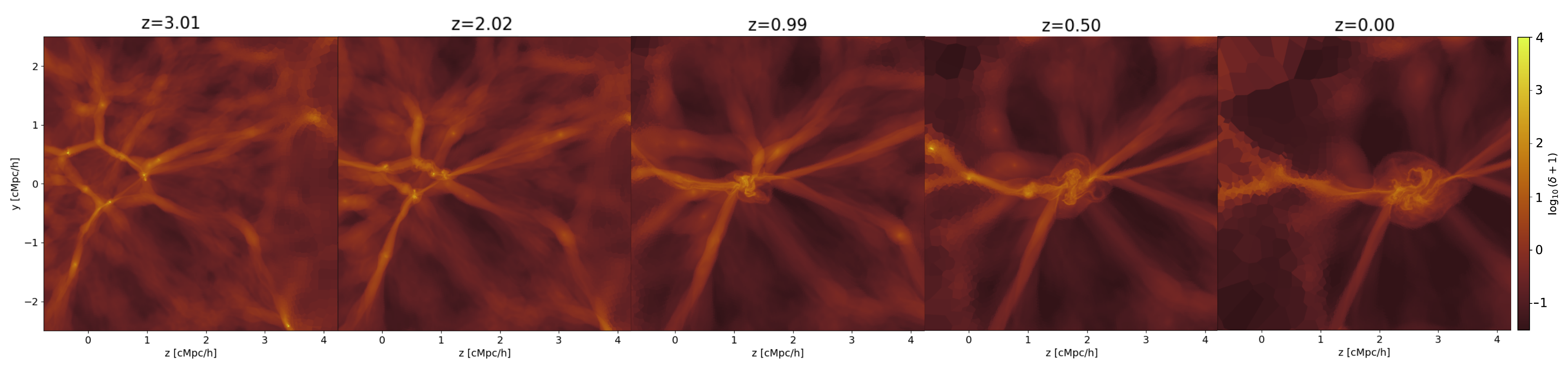}
\caption{From left to right: gas overdensity in a fixed
comoving region of $5\times5~\mathrm{cMpc}/h$ centred on the filament,
projected along the $x$ axis, at $z=3.01$, $2.02$, $0.99$, $0.50$, and $0.00$.}
\label{fig:assembly}
\end{figure*}

We validated the pipeline on a synthetic, incompressible Gaussian velocity field with a prescribed Kolmogorov spectrum. The direct-on-cells estimator recovers the non-intermittent scaling $\zeta_p/\zeta_3=p/3$ to high order, and returns a flatness of $5/3$ for the three-dimensional increment norm, the expected value for the magnitude of a Gaussian vector (a Maxwell distribution) rather than a signature of intermittency. Repeating the same measurement on a grid-interpolated version of the field reproduces the spurious high-order flattening described above, confirming its origin.

Because the inertial range spanned by a single filament is short, the individual exponents $\zeta_p$ are difficult to measure directly. We therefore use extended self-similarity \citep{1993PhRvE..48R..29Benzi}, plotting each $S_p(\ell)$ against $S_3(\ell)$ rather than against $\ell$. Intermittent flows follow clean power laws $S_p\propto S_3^{\,\zeta_p/\zeta_3}$ over a range far wider than the nominal inertial range, so that the relative exponents $\zeta_p/\zeta_3$ can be extracted robustly from the log--log slope. Because the extended-self-similarity power law extends well beyond the nominal inertial range, restricting the fit to that window would discard most of its advantage. We therefore fit these slopes over the full sampled separation range, retaining every bin that carries at least $2\times10^3$ pairs, and quote the fit uncertainties as error bars.

We assess the robustness of the $z=0$ exponents in two further ways, beyond the fit uncertainty quoted as error bars. First, we bootstrap the random large-separation pair sampling over $N_{\rm boot}=20$ independent draws. The resulting scatter in $\zeta_p/\zeta_3$ is comparable to the fit uncertainty across all orders, confirming that the exponents are limited by the intrinsic pair statistics rather than by the fitting procedure. Second, we split the filament into four equal-count sub-volumes along its long axis and measure the exponents independently in each. The four sub-volumes agree within their individual uncertainties up to sixth order, with the inter-region spread growing at the seventh and eighth orders where the reduced per-region statistics dominate. The $z=0$ scaling is therefore not driven by any single sub-region or merger event. 

The measured $\zeta_p/\zeta_3$ are compared with a set of reference models. The non-intermittent Kolmogorov prediction is the straight line $\zeta_p/\zeta_3=p/3$. Departures from it are described by the She--L\'ev\^eque family \citep{1994PhRvL..72..336SheLeveque,1994PhRvL..73..959Dubrulle},
\begin{equation}
\zeta_p = (1-\Delta)\,\frac{p}{3} + C\left[\,1-\left(1-\frac{\Delta}{C}\right)^{p/3}\,\right],
\label{eq:sheleveque}
\end{equation}
where $\Delta=2/3$ is fixed by the four-fifths law ($\zeta_3=1$) and $C$ is the codimension of the most dissipative structures. $C=2$ for the one-dimensional vortex filaments of the original incompressible model, and $C=1$ for two-dimensional sheets. The sheet-like limit is the one expected when dissipation is organised by shocks, as in the supersonic turbulence model of Boldyrev \citep{2002ApJ...569..841Boldyrev,2002ApJ...573..678Boldyrev,2013MNRAS.436.1245Federrath}. It is more intermittent than the filamentary case (lower $\zeta_p/\zeta_3$ at high order). As an extreme lower bound we also show the bifractal Burgers limit of a field dominated by shock discontinuities, $\zeta_p=\min(p,1)$, for which $\zeta_p/\zeta_3\to1$ at high order \citep{2007PhR...447....1BecKhanin}. 

We complement the exponents with the scale-dependent flatness $F(\ell)=S_4(\ell)/S_2(\ell)^2$, which equals $3$ for a Gaussian increment distribution and rises above it towards small separations whenever the field is intermittent. In the inertial range $F(\ell)\propto\ell^{\,\zeta_4-2\zeta_2}$, so that a flatness growing at small scales is the direct counterpart of exponents bending below $p/3$. 

\section{A filament assembled through hierarchical merging}
\label{sec:assembly}

In this first results section, we present the general properties of the simulated filament
at the present day, and we describe how it assembled over cosmic time.

In Fig.~\ref{fig:z=0}, we show the gas overdensity (left panel) and the gas
temperature (right panel) of the filament at $z=0$, projected along the $z$
axis. The filament connects two haloes of $M_{200}\simeq1.7$ and
$1.2\times10^{12}\,M_\odot$ that mark the secondary nodes of the cosmic web,
over a node-to-node length of $\simeq5.5~\mathrm{cMpc}/h$ ($\simeq8~\mathrm{Mpc}$),
and that it extends well beyond them into the surrounding intergalactic medium.
Its diffuse gas and dark matter overdensity falls from $\langle\delta\rangle\simeq12$
near the spine to $\simeq1$ by a transverse radius of $1~\mathrm{cMpc}/h$, with a
linear mass density $\lambda\simeq5\times10^{11}\,M_\odot\,(\mathrm{cMpc}/h)^{-1}$. With a median gas cell size of
$\simeq8~\mathrm{kpc}$ (see Fig.~\ref{fig:hists}), the simulation resolves the
internal structure of the diffuse gas well below the transverse width of the
filament. The overdensity map reveals a dense, cool, and relatively narrow spine,
around which the gas is markedly clumpy and filamentary rather than smooth. 

In the temperature map, we observe that away from this spine the diffuse gas is
shock-heated into the warm-hot regime, at $10^5$--$10^7~\mathrm{K}$, which fills
the body of the filament (see the phase-space diagram in Fig.~\ref{fig:phase}).
The sample is not temperature-selected, so it also retains the cold photoionised
component present at lower overdensity (Sect.~\ref{sec:selection}). This hot and
structured envelope already shows that the intergalactic gas of the filament is far
from a quiescent, laminar inflow. We also note, in the temperature map, thin, cold
and dense structures that thread the hot interior, morphologically reminiscent of
the cold streams studied at higher redshift \citep{2019MNRAS.484.1100Mandelker}.
Their survival, origin, and fate in a $z=0$ filament are beyond the scope of this
first paper and motivate a dedicated study.

In Fig.~\ref{fig:assembly}, we show the assembly history of the filament,
through the gas overdensity projected along the $x$ axis at $z=3.01$, $2.02$,
$0.99$, $0.50$, and $0.00$ (from left to right). The assembly is
not laminar. Rather than growing by the steady accretion of matter onto a single
strand, the filament is woven from several converging proto-filaments. At early
times (left panels), we observe that the region hosts several distinct strands
that accrete their own surroundings. As the system evolves, these strands merge
hierarchically into a single spine, a process that is punctuated by the infall
and coalescence of the haloes that mark the secondary nodes.

This mode of growth matters for what follows. By assembling from several
colliding streams rather than from a single ordered inflow, the filament injects
kinetic energy into its diffuse gas across a wide range of scales at once,
through accretion, through mergers, and through the shear between the converging
strands. It thus sets the stage for the turbulence that we characterise in the
rest of this work. In the following, we examine this process at the epoch when
the assembly is most active.

\section{Vorticity generation at accretion shocks}
\label{sec:vort}

\begin{figure*}
     \centering
     \includegraphics[width=\linewidth]{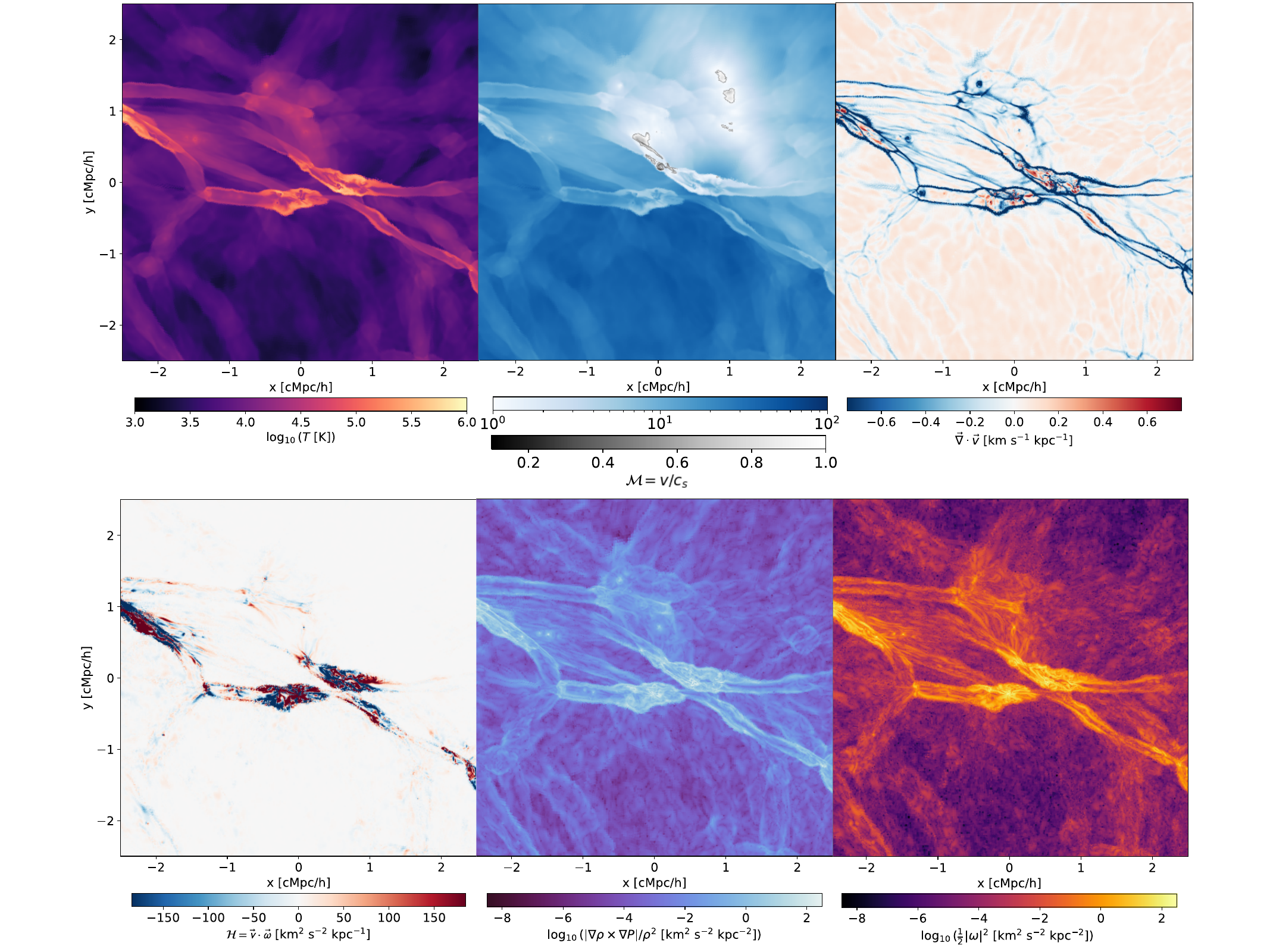}
\caption{Slices of physical quantities on $5\times5~\mathrm{cMpc}/h$
through the $x$--$y$ plane at $z\simeq2$, without halo excision. Top row: gas temperature (left), Mach number (centre) and velocity divergence
(right). Bottom row: kinetic helicity (left),
baroclinic term (centre) and enstrophy (right). Axes are in comoving units.}
\label{fig:vort_z2}
\end{figure*}

We now investigate how the diffuse gas acquires its rotational
motions during the assembly of the filament. We focus on the epoch $z\simeq2$,
when the filament is in its most active assembly phase, with the proto-filaments
converging and accreting the surrounding gas.

In Fig.~\ref{fig:vort_z2}, we show several quantities in a thin slice of $5\times5~\mathrm{cMpc}/h$
through this forming structure, taken in the $x$--$y$ plane at a height of
$0.55~\mathrm{cMpc}/h$ above the filament centre. All the gas is shown here,
without the halo excision applied to the quantitative analysis
(Sect.~\ref{sec:selection}), so that the accretion flow and its shocks are
visible in full.

In the top row, we present the quantities that characterise the compressive
state of the flow. The temperature map (left) separates the
hot interior of the forming filament and its nodes from the cold gas of the
surrounding sheets and voids. In the Mach number map (centre), this cold, infalling gas is highly supersonic ($\mathcal{M}>1$),
while the hot, pressure-supported interior of the filament is subsonic
($\mathcal{M}<1$), so that the accretion delivers the gas onto the filament at
supersonic speeds. The velocity divergence (right) then
locates where this flow is decelerated. Its strongly converging
surfaces ($\nabla\cdot\boldsymbol{v}<0$) trace the accretion shocks that bound
the filament, together with the interfaces between the several proto-filaments
converging onto the spine, and that they are strongest around the two main haloes within the filament, where the accretion is most intense.

The gas does not, however, reach these shocks head-on. The assembling strands carry their own rotation, an angular momentum acquired during their anisotropic gravitational collapse. This rotation reflects the vorticity of the cosmic velocity field, which numerical work
finds to wind coherently around filaments in a quadrupolar pattern \citep{2015MNRAS.446.2744Laigle}, and cosmic filaments are indeed observed to rotate \citep{2021NatAs...5..839Wang}.

We trace this rotation with the kinetic helicity
(bottom row, left panel of Fig.~\ref{fig:vort_z2}). The helicity is enhanced along the spine and changes sign coherently across it, which is the signature of an ordered,
large-scale rotation of the strands about the filament axis. As a consequence,
the accreting gas spirals onto the filament and meets the shock-bounded
interfaces obliquely, rather than along the shock normal.

At such an oblique shock, the density and pressure gradients are no longer
aligned, and vorticity is generated through the baroclinic effect. In the bottom row of Fig.~\ref{fig:vort_z2}, we compare the baroclinic term (middle) and the enstrophy (right) to show that this is where the rotational motions are seeded. We observe that both are enhanced along
precisely the strand interfaces traced by the shocks in the top right panel of Fig.~\ref{fig:vort_z2}, and that both
fall to negligible values in the voids. This close spatial correspondence
between the baroclinic source and the enstrophy shows that the oblique arrival of
the accreted gas at these shocks, which misaligns the pressure and density
gradients, is the likely mechanism that seeds the vorticity preceding turbulence in the diffuse gas. We examine its temporal counterpart in
Sect.~\ref{sec:growth}.

Baroclinic vorticity generation at accretion shocks is well established in
clusters and in the large-scale web
\citep{2008Sci...320..909Ryu,2017MNRAS.464..210Vazza,2017MNRAS.464.4448Wittor}. It is distinct from the kinematic vorticity of the collisionless velocity
field, which arises at caustics through shell crossing
\citep{1999A&A...343..663Pichon} and sets the coherent, large-scale winding
rather than the small-scale vorticity that feeds the cascade. What is
specific to a filament is that its collapse proceeds along two axes rather
than three, which leaves the strands rotating about the filament axis. The gas here therefore meets the shock obliquely by construction rather than incidentally, and our targeted zoom resolves this mechanism directly within the diffuse gas, rather than inferring it from the denser, collapsed regions.

\section{The growth of turbulence over cosmic time}
\label{sec:growth}

\begin{figure}
     \centering
     \includegraphics[trim=0 0 0 0,clip,width=\linewidth]{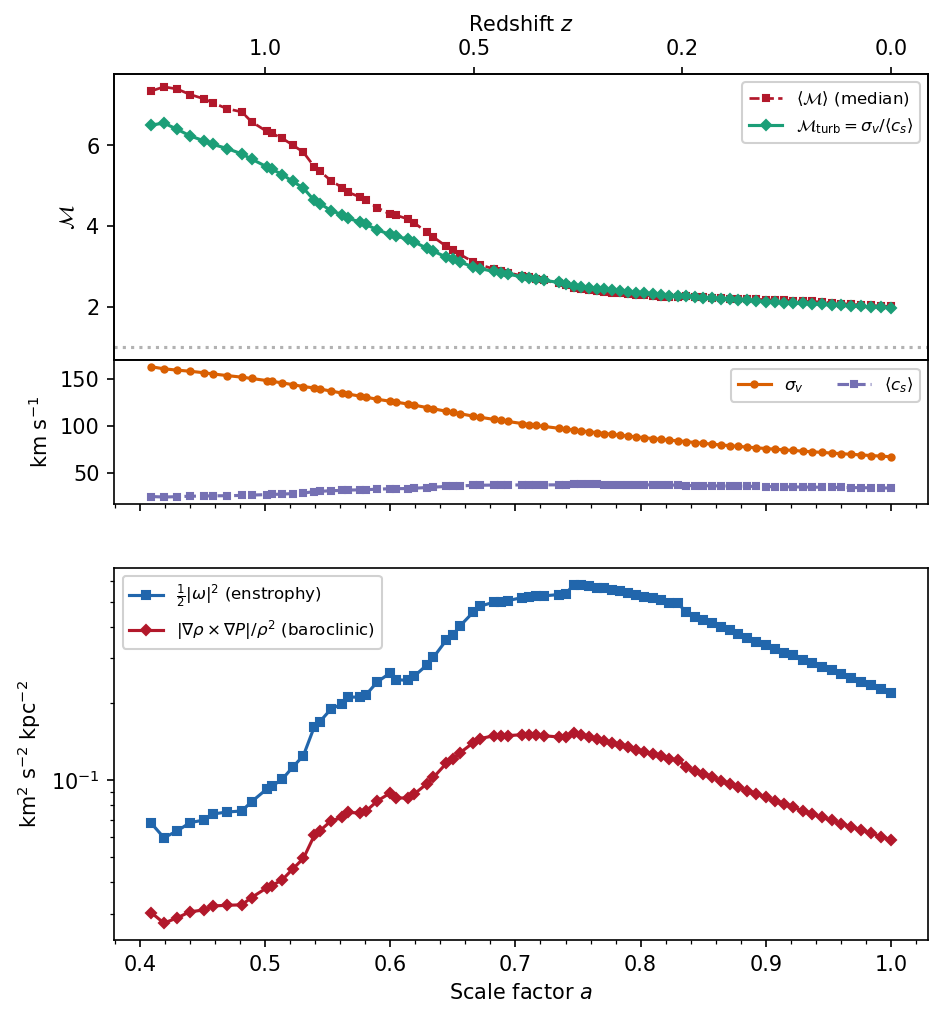}
\caption{Growth of turbulence in the diffuse filament gas over cosmic time,
against scale factor $a$ (bottom) and redshift $z$ (top). Top:
cell-by-cell Mach number (red) and turbulent Mach number
$\mathcal{M}_{\rm turb}=\sigma_v/\langle c_s\rangle$ (green). The velocity
dispersion $\sigma_v$ (orange) and mean sound speed $\langle c_s\rangle$ (blue) are shown in the associated sub-panel. Bottom: median enstrophy (blue) and baroclinic
term (red), peaking together near $a\simeq0.72$.}
\label{fig:time_evo}
\end{figure}

In this section, we follow the diffuse gas across cosmic time, in order to see
how the vorticity that is locally seeded at the accretion shocks develops into
the turbulent state that we observe at the present day.

In Fig.~\ref{fig:time_evo}, we show the evolution of the turbulence diagnostics
of the retained, non-star-forming gas, from $a=0.41$ ($z\simeq1.4$) to $z=0$,
computed on the identical selection at every snapshot (Sect.~\ref{sec:selection}).
Because the halo excision leaves a few rare, high-vorticity cells that would
dominate an arithmetic mean, we read the rotational diagnostics and the
cell-by-cell Mach number through the median of the retained gas rather than the
mean (Sect.~\ref{sec:diag}).

In the top panel of Fig.~\ref{fig:time_evo}, we present the evolution of the
kinematics of the gas as the assembly slows down. The
three-dimensional velocity dispersion of the retained gas falls from
$\sigma_v\simeq160~\mathrm{km\,s^{-1}}$ at early times to
$\simeq67~\mathrm{km\,s^{-1}}$ at $z=0$, while the mean sound speed rises and then
plateaus as the gas is heated into the warm-hot regime. As a result, the
turbulent Mach number $\mathcal{M}_{\rm turb}=\sigma_v/\langle c_s\rangle$ drops
from $\simeq6.5$, deep in the supersonic accretion regime, towards the sonic
scale, reaching $\simeq2$ at the present day. A comparison with the cell-by-cell
Mach number is instructive. At early times the latter lies well
above $\mathcal{M}_{\rm turb}$, because its median still carries the coherent,
large-scale infall on top of the turbulent motions. The two Mach numbers converge
by $z\simeq0.45$ and remain equal thereafter, which we interpret as the epoch at
which the ordered accretion has subsided and the velocity field of the diffuse
gas has become genuinely turbulent. We note that the present-day
$\mathcal{M}_{\rm turb}\simeq2$ remains mildly supersonic, which already implies a
substantial non-thermal pressure support that we quantify in
Sect.~\ref{sec:implications}.

In the bottom panel of Fig.~\ref{fig:time_evo}, we show the evolution of the
median enstrophy and baroclinic term, which we plot on a common axis by virtue of
their shared dimension of an inverse time squared. Both rise
through the assembly phase and peak together near $a\simeq0.72$ ($z\simeq0.4$)
before subsiding. This joint peak is the temporal counterpart of the spatial
co-location of the two quantities that we observed at $z\simeq2$
(Fig.~\ref{fig:vort_z2}), and it places the epoch of most active baroclinic
injection squarely within the shock-dominated assembly phase. Beyond the peak, we
observe that the baroclinic term weakens markedly while the enstrophy subsides
only mildly. 

The term-by-term enstrophy budget confirms this reading (Appendix~\ref{app:enstrophy}). The vorticity is injected at the accretion shocks during the assembly, where the baroclinic term generates it and compression amplifies it, and is subsequently sustained by the vortex-stretching of the developed cascade rather than continuously regenerated by the baroclinic term, whose contribution fades away from the shocks.

The statistical character of these motions changes in step with the flow. In the
following section, we examine it directly, through the scaling of the velocity
increments.

\section{From shock-dominated intermittency to a developed cascade}
\label{sec:cascade}

\begin{figure}
     \centering
     \includegraphics[trim=0 0 1250 0,clip,width=\linewidth]{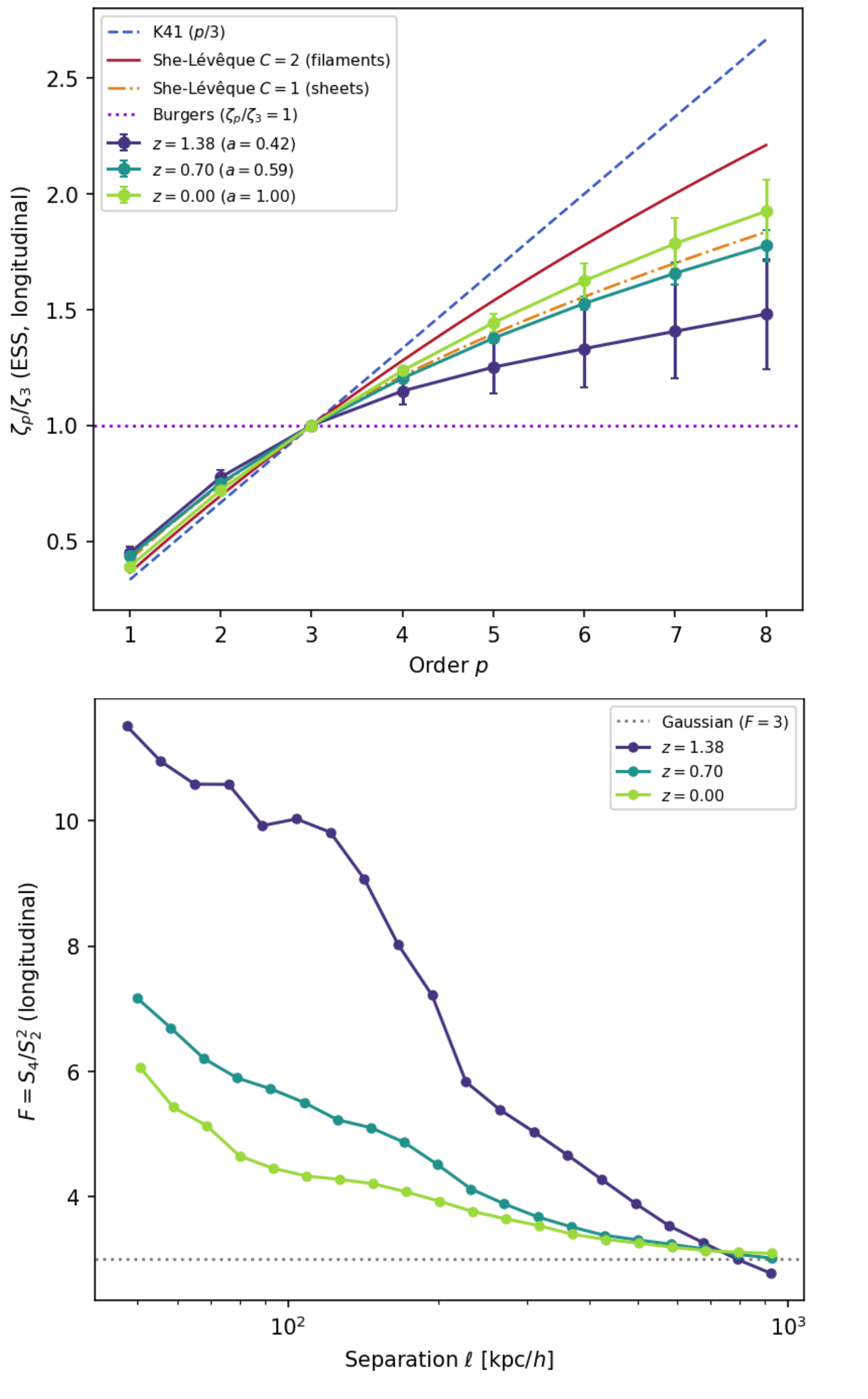}
\caption{Top: relative exponents
$\zeta_p/\zeta_3$ from extended self-similarity, compared with the Kolmogorov
($p/3$, dashed blue), She--L\'ev\^eque ($C=2$ filaments in solid red, $C=1$ sheets in dash-dotted orange) and bifractal Burgers (dotted blue)
references. Error bars are the fit uncertainties. Bottom: longitudinal flatness
$F=S_4/S_2^2$ against comoving separation $\ell$, with the Gaussian value $F=3$.}
\label{fig:interm}
\end{figure}

In this final results section, we quantify the intermittency of the turbulent cascade and
follow its evolution, through the scaling of the velocity increments
$\delta v_\parallel(\ell)$, that is, the velocity differences between pairs of
cells a comoving distance $\ell$ apart.

Fully developed Kolmogorov turbulence predicts increments whose statistics are
self-similar across scales. Intermittency is the departure from this baseline,
namely the excess of rare, extreme increments produced by the shocks and shear
layers where the energy dissipates, which populate the tails of the increment
distribution. Because raising the increment to a higher power $p$ weights the
largest values ever more heavily, the longitudinal structure functions
$S_p(\ell)=\langle|\delta v_\parallel|^p\rangle$ probe progressively deeper into
these tails. In Fig.~\ref{fig:vsf}, we show that, computed directly on the
Voronoi cells (Sect.~\ref{sec:vsf}), they follow clean power laws over the fit
range $[80,700]~\mathrm{ckpc}/h$ at every order from $p=1$ to $8$, so that the
diffuse gas sustains a well-defined inertial range. We then measure the
intermittency of the cascade through the relative exponents $\zeta_p/\zeta_3$
that we extract from these functions by extended self-similarity.

In the top panel of Fig.~\ref{fig:interm}, we present the relative exponents at three epochs,
$z=1.38$, $0.70$, and $0.00$, together with the reference models of developed and
shock-dominated turbulence. The exponents evolve between these
models. During the assembly phase ($z=1.38$), when the gas is strongly
supersonic ($\mathcal{M}_{\rm turb}\simeq6.5$) and its dynamics dominated by
accretion shocks, the exponents are the flattest and most intermittent of the
three epochs. They lie well below the sheet-like She--L\'ev\^eque form ($C=1$)
and approach the bifractal Burgers limit $\zeta_p/\zeta_3\to1$ of a purely
shock-dominated field. By the present day, we observe that they have risen away
from that limit to track the $C=1$ prediction across the whole range of orders,
which is the hallmark of a developed multifractal cascade. To place this visual
reading on a quantitative footing, we compare the $z=0$ exponents with each
reference model through a $\chi^2$ over the orders up to six, which should be read as a relative preference rather than an absolute measure of goodness of fit.
We find that the sheet-like $C=1$ form is favoured, with $\chi^2/\mathrm{dof}
\simeq9.6$, ahead of the filamentary $C=2$ form ($\simeq14.6$) and far ahead of
the non-intermittent Kolmogorov ($\simeq71.0$) and bifractal Burgers ($\gg100$)
limits.

In the bottom panel of Fig.~\ref{fig:interm}, we show the scale-dependent flatness
$F(\ell)=S_4/S_2^2$, the normalised fourth moment that gauges how heavy the wings
of the increment distribution are relative to a Gaussian. It tells the same story
independently. At every epoch it rises above the Gaussian value
$F=3$ towards small separations, which is the defining feature of intermittency,
but that it is highest during the shock-dominated assembly phase and decreases
towards the present day, continuously across all snapshots (see
Appendix~\ref{app:timeinterm}). This decline is real, yet it marks not a
weakening of the turbulence but its maturation. A shock-dominated field is
trivially intermittent, its high flatness reflecting the sharp, near-binary
contrast between a few extreme discontinuities and the smooth flow between them.
As the supersonic accretion subsides and the turbulent Mach number falls towards
the sonic scale, this near-bifractal field gives way to a developed multifractal
cascade, in which a continuous distribution of singularity strengths replaces
that sharp contrast. The flatness falls because the extreme, shock-driven tails
recede, even as the cascade that organises the intermittency becomes fully
developed. What emerges over cosmic time is thus not intermittency itself, but
the developed cascade that structures it.

At $z=0$, this cascade is characterised by sheet-like dissipative structures,
consistent with the shocks and shear layers from which the vorticity was
originally seeded (Sect.~\ref{sec:vort}). We find that it is close to
statistically isotropic at second order, and we examine the transverse exponents,
with the caveats they carry, in Appendix~\ref{app:iso}. The cascade also carries no robust compressive signature at $z=0$. Its high-order exponents rise well above the bifractal Burgers value (Fig.~\ref{fig:interm}), so the velocity field follows a developed multifractal rather than the near-bifractal scaling that strong compressibility would imprint. The third-order exponent, measured directly from $S_3(\ell)$ rather than through extended self-similarity, is compatible with the incompressible value, $\zeta_3\simeq0.9$--$1.0$ across the plausible fit windows. On its own it is only a weak diagnostic, since both the incompressible and the Burgers limits share $\zeta_3=1$, but it confirms that the four-fifths anchor underlying our relative exponents holds. A direct measurement of the compressible energy transfer, through the density-weighted variable $\rho^{1/3}\boldsymbol{v}$ \citep{2007ApJ...665..416Kritsuk} or the exact compressible flux law \citep{2011PhRvL.107m4501Galtier}, is left to future work.

\section{Discussion}
\label{sec:discussion}

By targeting a single filament and following it to the present day, this first
WEFT simulation resolves a sequence that large-volume simulations cannot trace
and cluster-centred zooms only glimpse. The diffuse gas of a cosmic filament is
not a still reservoir but the seat of a turbulent cascade that assembles over
cosmic time. Vorticity is seeded baroclinically at the oblique, shock-bounded
interfaces where the rotating proto-filaments accrete their surroundings, and the flow evolves from a
supersonic, shock-dominated state, statistically close to the bifractal Burgers
limit, into a fully developed multifractal cascade whose longitudinal exponents
lie closest to the sheet-like She--L\'ev\^eque prediction by $z=0$. That this
sheet-dominated intermittency emerges in a purely hydrodynamic calculation is itself informative. The dissipative structures are set by shocks and shear layers alone, with no magnetic field required to organise them, and the cascade does not inherit the geometry of its host either. A quasi-one-dimensional filament might be expected to dissipate in filamentary structures, yet the exponents follow the sheet-like form, so the dissipative geometry is set by the flow rather than by the container.

\subsection{Implications for the baryon budget and the observability of the WHIM}
\label{sec:implications}

This turbulent state carries direct consequences for the baryon budget of filaments. Today, the diffuse gas is mildly supersonic, with
$\mathcal{M}_{\rm turb}\simeq2$, so that the turbulent pressure is about twice the thermal pressure of the diffuse gas. This support is not inflated by ordered motion. The coherent
rotation about the spine seen in the helicity field contributes only
$\simeq2\%$ of the velocity variance $\sigma_v^2$ that sets $P_{\rm turb}$, and all coherent flows together at most $\simeq7\%$, so that subtracting them lowers the ratio only from $2.0$ to $1.9$. The factor of two is therefore a robust measurement, not an upper limit. Any attempt to weigh the WHIM or infer its thermodynamic state from pressure-based observables must therefore account for
this non-thermal component, much as turbulent pressure already complicates the hydrostatic mass estimates of galaxy
clusters \citep{2026arXiv260700610Lebeau,2026A&A...707A.336Lebeau}.

The same motions set the density and velocity fluctuations that shape the
observable signatures of the missing baryons, from the line widths and centroid
shifts of X-ray WHIM lines \citep{tanimura2022xray,migkas2025detection} and the
statistics of H\,{\sc i} 21~cm
emission \citep{2017MNRAS.468..857Kooistra,2019MNRAS.490.1415Kooistra} to the
absorption imprint in the Lyman-$\alpha$
forest \citep{2006A&A...445..827Richter} and the power spectrum and higher-order statistics measured by line-intensity mapping \citep{Kovetz2017LIM}. In the coming SKA and X-ray microcalorimetry era \citep{Pan2026SKA,Cuciti2026SKA}, such forecasts can be grounded in a self-consistent turbulent state rather than an imposed prescription.

A further, still largely unexplored consequence concerns the non-thermal
particle content of filaments. Turbulence can reaccelerate relativistic
electrons through second-order Fermi processes \citep{BrunettiVazza2020},
powering the diffuse synchrotron emission now detected in stacked filaments and
intercluster bridges \citep{Vernstrom2021,Vernstrom2023,Govoni2019}. Our purely
hydrodynamic calculation does not follow the magnetic fields needed to predict
this emission directly, but it supplies precisely the turbulent input, that is, its
amplitude, intermittency, and scaling, on which such models depend.

\subsection{Comparison with previous works}
\label{sec:comparison}

Placed in the wider context of the cosmic web, this transition from a
shock-dominated flow to a developed cascade mirrors, at lower density and
temperature, the gas motions now being resolved in the intracluster medium by
X-ray spectroscopy \citep{XRISM2025} and by cosmological
simulations \citep{2025A&A...704A..14Lebeau}, identifying filaments as the
diffuse extension of a single picture in which structure formation sets the
baryons in motion on every scale. This same medium sets the boundary condition
for the galaxies that grow within the filament and fall along it towards
clusters, an issue that this targeted, high-resolution view is well suited to
address.

The sheet-like ($C=1$) scaling we measure is also the prediction for magnetised
turbulence, in which dissipation is organised into current
sheets \citep{2000PhRvL..84..475Muller,2010ApJ...720..742Kowal}. That the same
scaling arises here from shocks and shear layers alone, with no magnetic field,
suggests that the sheet-dominated character of the cascade is a robust feature
of the diffuse web rather than a signature of any single organising mechanism.

\subsection{Limitations and perspectives}
\label{sec:limitations}

The physical sequence we trace rests on mechanisms generic to filament assembly
and should hold across the population. 

The filament we simulate is a typical member of that population
(Sect.~\ref{sec:assembly}): its length lies close to the median of the TNG300 filaments of \citet{galarraga2020populations} and within their short class, while the two group-scale nodes it connects sit below their mean node mass, placing it among the lower-mass, more tenuous filaments rather than the dense cluster-connecting bridges. The numerical values we report, from the present-day turbulent Mach number to the amplitude of the rotational diagnostics, are specific to this pathfinder run and subject to cosmic variance. A representative sample of filaments will be needed to turn the former into a population-level statement and to calibrate the latter, which is the aim of the
WEFT programme.

The calculation is purely hydrodynamic, and the magnetic fields threading real
filaments would add a non-thermal pressure of their own and could modify the
intermittency of the cascade. Reassuringly, the sheet-like ($C=1$) scaling we
measure coincides with the prediction for magnetised turbulence
\citep{2000PhRvL..84..475Muller,2010ApJ...720..742Kowal}. Including magnetic
fields would therefore plausibly preserve the sheet-dominated character of the
cascade, shifting its physical origin from shocks and shear layers to current
sheets while quantitatively modifying the exponents.

Three questions follow and motivate the continuation of the WEFT programme. The
first is the full multiscale structure of the cascade, the distribution of
singularity strengths that a bifractal or lognormal description cannot capture,
which we will address through a wavelet-based multifractal analysis. The second
is how the vorticity seeded at the shock-bounded interfaces spreads inward to
fill the diffuse core, with the hierarchical merging of the proto-filaments and
the growth of Kelvin--Helmholtz instabilities along the sheared interfaces as
natural candidate agents. The third is what the enriched, multiphase gas would add. A future
run incorporating metal enrichment, a star-forming interstellar medium, and the
associated feedback would open the cold gas of galaxies and their circumgalactic
medium to the [C\,{\sc ii}] and CO tracers of line-intensity
mapping \citep{Kovetz2017LIM}, connecting the diffuse reservoir characterised
here to the galaxies that grow within it.

\section{Conclusions}
\label{sec:conclusion}

In this first paper of the WEFT project, we have turned the cosmological
zoom-in technique on a single cosmic filament and followed it with the
moving-mesh code \texttt{AREPO} from $z=63$ to $z=0$, reaching a median gas
cell size of $\simeq8~\mathrm{kpc}$ within it. To our knowledge this is the
first cosmological simulation to resolve the internal gas dynamics of a
filament as its primary target down to the present day. We characterised the
emergence of turbulence in its diffuse gas through its thermodynamics,
its vorticity content, and the scaling of its velocity increments. Our main
results are the following.

\begin{enumerate}
\item The filament assembles by the hierarchical merging of several converging
proto-filaments rather than by laminar accretion, injecting kinetic energy into
the diffuse gas across a wide range of scales.

\item Vorticity is seeded baroclinically at the accretion shocks of the
assembly epoch, where the gas spiralling onto the rotating strands reaches the
shock-bounded interfaces obliquely. The baroclinic term and the enstrophy peak
together, both spatially at $z\simeq2$ and temporally near $a\simeq0.72$.

\item The turbulent Mach number falls from $\simeq6.5$ during the supersonic
assembly phase to $\simeq2$ at $z=0$. Ordered accretion gives way to genuinely
turbulent motions by $z\simeq0.45$, when the turbulent and cell-by-cell Mach
numbers converge.

\item The velocity field evolves from a shock-dominated, near-bifractal state,
statistically close to the Burgers limit, into a developed multifractal cascade
whose longitudinal exponents lie closest to the sheet-like She--L\'ev\^eque
model ($C=1$) at $z=0$. The cascade is close to isotropic at second order and
carries no robust compressive signature, with $\zeta_3\simeq1$. Its turbulent
pressure reaches about twice the thermal pressure of the diffuse gas.
\end{enumerate}

Turbulence is thus an intrinsic and quantifiable property of the diffuse gas of
cosmic filaments. This pathfinder run supplies the turbulent input, its
amplitude, intermittency, and scaling, on which forecasts of the observability of
the WHIM can be based, and it sets the stage for the continuation of the WEFT
programme towards a representative sample of filaments and the full multiscale
characterisation of their turbulence.

\begin{acknowledgements}
We would like to thank Volker Springel for providing access to the development version of the \texttt{AREPO} code. We also acknowledge support by the Israel Science Foundation (grant no. 1388/24), and thank the Center for Information Technology of the University of Groningen for their support and for providing access to the Hábrók high performance computing cluster. TL also thanks colleagues at the Kapteyn Astronomical Institute, in particular Guillaume Desprez, for helpful discussions and comments. During the preparation of this manuscript, the authors used a large language model (Claude, Anthropic) to assist with language editing and condensing the text. All scientific content, analysis, and interpretation are the authors' own, and the authors take full responsibility for the content, including all cited references.
\end{acknowledgements}

\section*{Data availability}
The raw simulation snapshots are available from the corresponding author on reasonable request owing to their volume. The analysis codes used in this study are available at \url{https://github.com/theolebeau/WEFT-project.git}. The \texttt{AREPO} code is available in its public version at \url{https://arepo-code.org/}, and in its development version upon request to the developers of the code.

\bibliographystyle{aa}
\bibliography{bibliography}

\begin{appendix}

\section{Resolution convergence}
\label{app:conv}

\begin{figure}[h]
     \centering
     \includegraphics[trim=0 0 1250 50,clip, width=\linewidth]{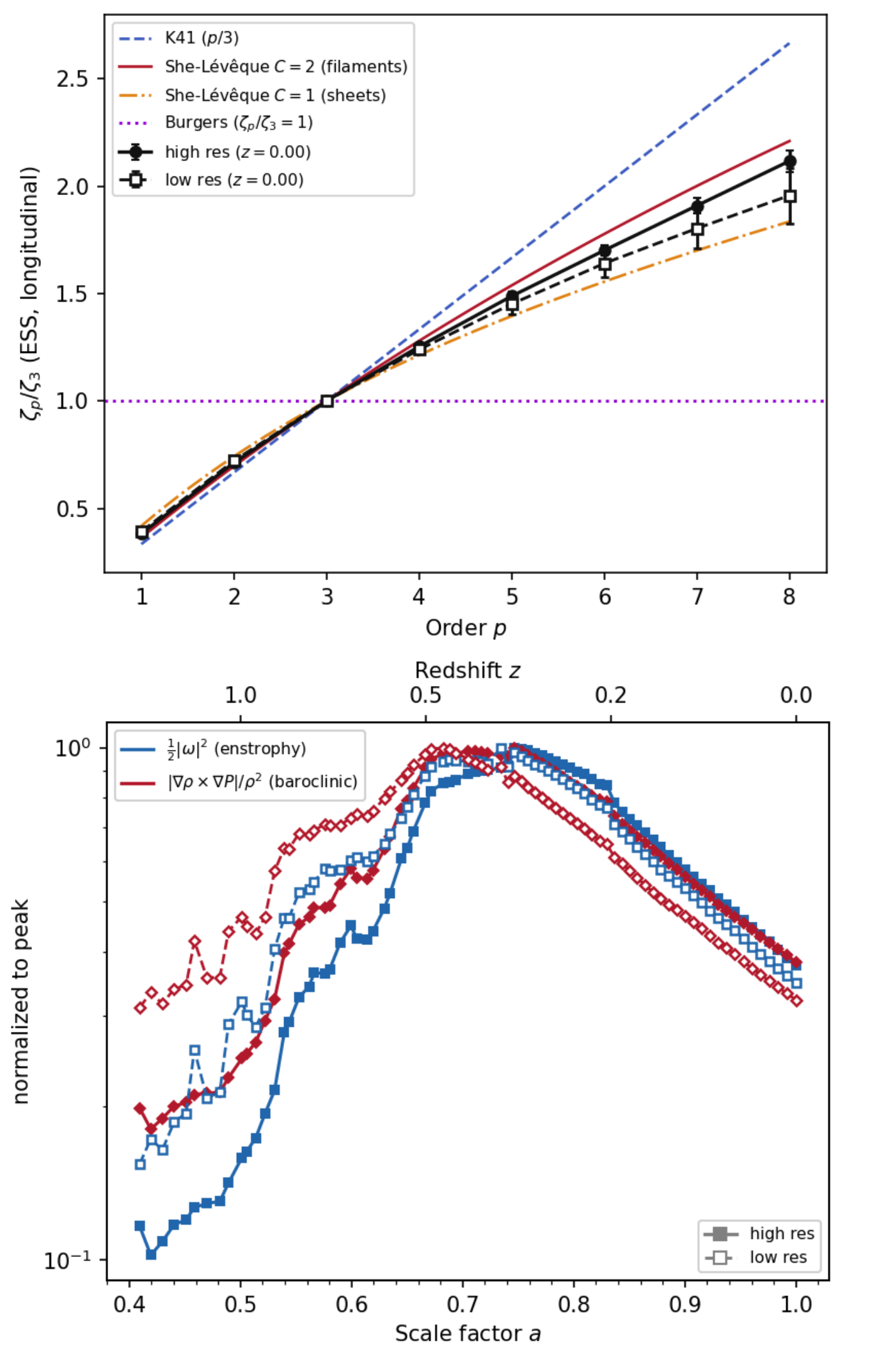}
\caption{Top: relative exponents $\zeta_p/\zeta_3$ at $z=0$ for the high- (filled circles) and low-resolution (open squares) runs, with the Kolmogorov (dashed blue), She--L\'ev\^eque ($C=2$ in solid red, $C=1$ in dashed-dotted orange) and Burgers (dotted blue) references. Error bars are the extended-self-similarity fit uncertainties. Bottom: evolution of the median enstrophy and baroclinic term against scale factor $a$, each normalised to its own peak.}
\label{fig:conv}
\end{figure}

We verify that the intermittency and vorticity diagnostics are not resolution
artefacts by comparing the primary run (effective resolution of $4096^3$ cells, labelled high res) with a lower-resolution
version (effective resolution of $2048^3$ cells, labelled low res) evolved from identical initial conditions with the same
physics (Fig.~\ref{fig:conv}). The relative scaling exponents $\zeta_p/\zeta_3$
measured at $z=0$ (top panel) agree within their fit uncertainties up to sixth order and both
track the sheet-like $C=1$ model. The enstrophy and baroclinic terms (bottom panel) are
gradient-based, so their absolute amplitude rises with resolution and is not
expected to converge, but their temporal morphology does, including the joint
peak near $a\simeq0.72$. During the early assembly phase ($a\lesssim0.6$) a
mild resolution-dependent offset remains in the relative onset of the
diagnostics, which is why this phase is described only qualitatively and is not
used in the quantitative analysis.

\section{Thermodynamic phase-space diagram}
\label{app:phase-space}

\begin{figure}[h]
     \centering
     \includegraphics[trim=0 0 650 0,clip, width=\linewidth]{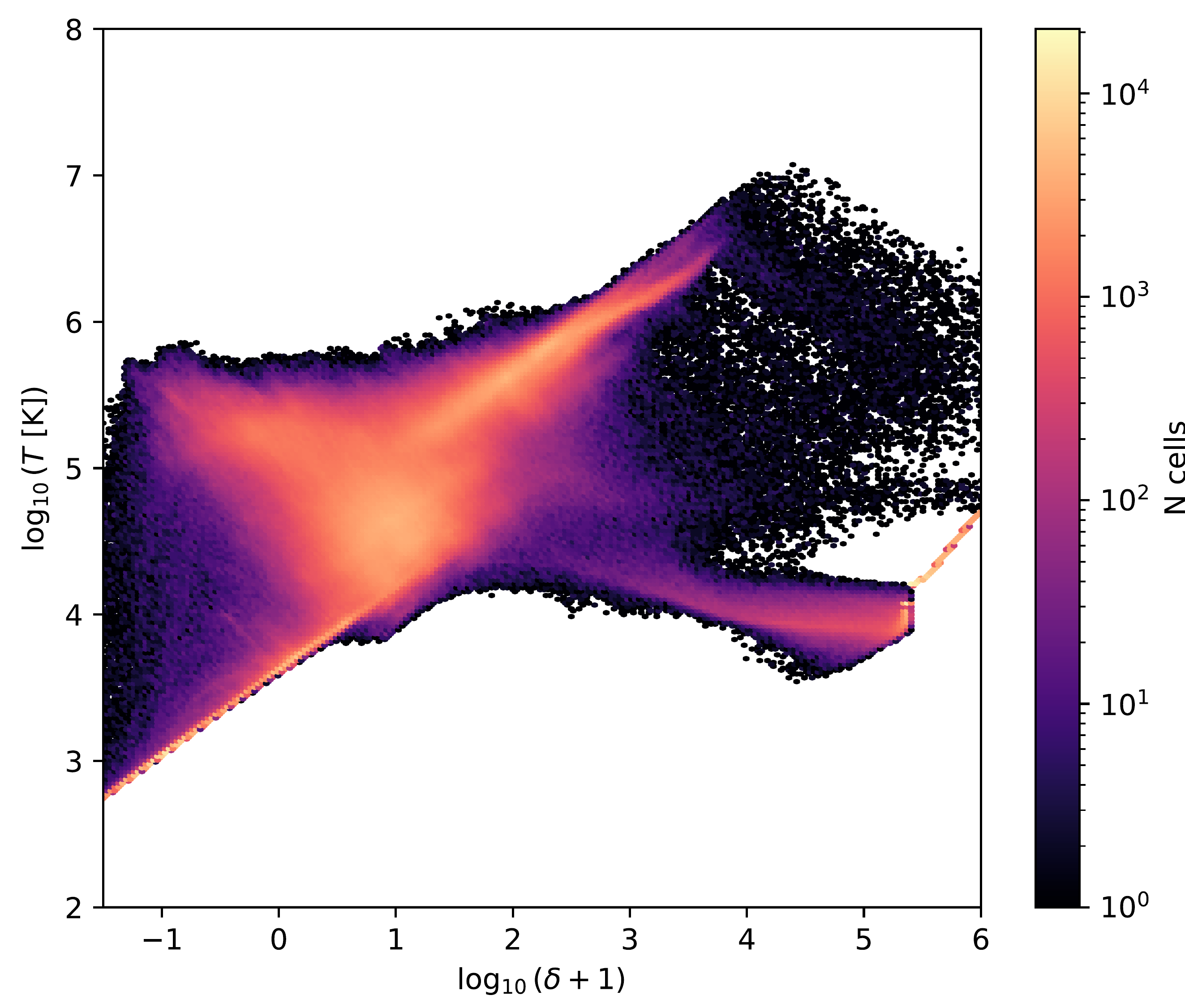}
\caption{Mass-weighted distribution of the filament gas in the overdensity--temperature plane at $z=0$. The orange dashed line marks the effective equation of state of \citet{2003MNRAS.339..289SpringelHernquist}.}
\label{fig:phase}
\end{figure}

Figure~\ref{fig:phase} shows the mass-weighted distribution of the filament gas in the overdensity--temperature plane at $z=0$. Three components stand out: a cold, diffuse phase at low overdensity, held near $10^4~\mathrm{K}$ by the photoionising background, a warm-hot phase at $10^5$--$10^7~\mathrm{K}$ that hosts the WHIM targeted in this work, and a star-forming branch lying on the effective equation of state of \citet{2003MNRAS.339..289SpringelHernquist}. The turbulence analysis retains the diffuse gas, both the cold and warm-hot phase and excises the star-forming branch, whose temperature and velocity follow the sub-grid model rather than resolved dynamics (Sect.~\ref{sec:selection}).

\section{Isotropy and transverse structure functions}
\label{app:iso}

\begin{figure}[h]
     \centering
     \includegraphics[width=0.8\linewidth]{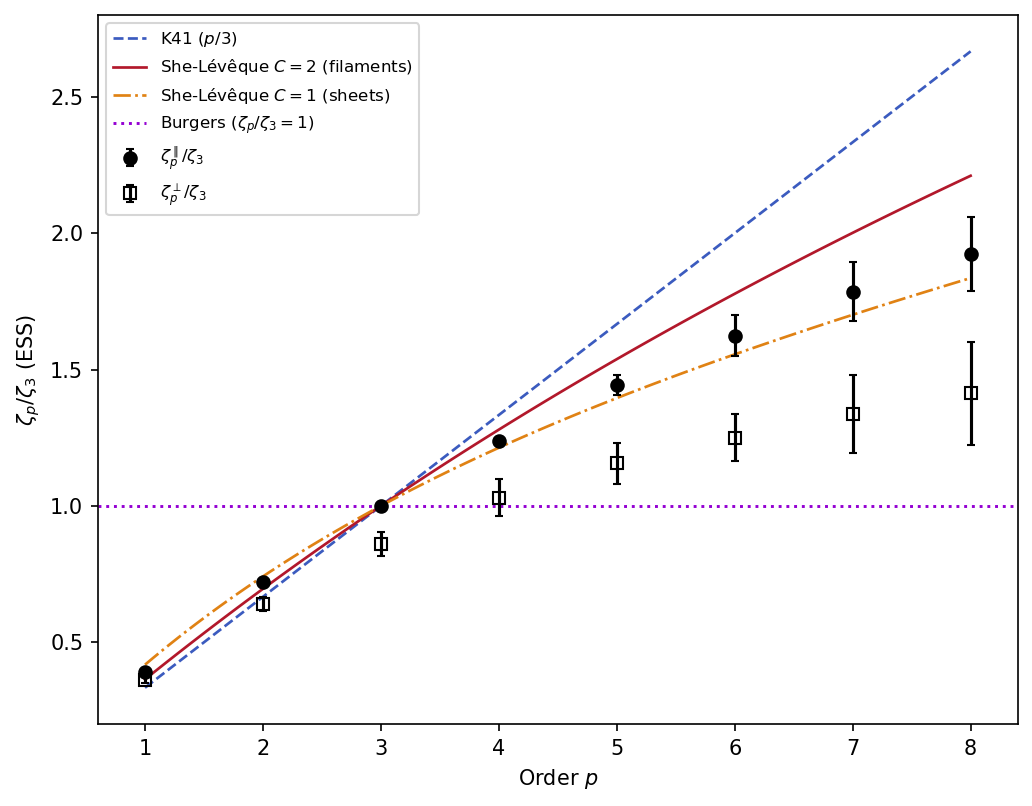}
\caption{Extended self-similarity and isotropy of the cascade at $z=0$. Relative
exponents $\zeta_p/\zeta_3$ of the longitudinal (filled circles) and transverse
(open squares) increments, both regressed against the same longitudinal $S_3$,
compared with the Kolmogorov, She--L\'ev\^eque ($C=1$, $C=2$) and Burgers
references. Error bars are the
extended-self-similarity fit uncertainties.}
\label{fig:iso}
\end{figure}

We assess the isotropy of the cascade by measuring, alongside the longitudinal
functions, the transverse structure functions $S_p^\perp(\ell)=\langle|\delta\boldsymbol{v}_\perp|^p\rangle$, where $\delta\boldsymbol{v}_\perp=\delta\boldsymbol{v}-(\delta\boldsymbol{v}\cdot\hat{\boldsymbol{\ell}})\,\hat{\boldsymbol{\ell}}$ is the full velocity increment in the plane perpendicular to the separation, so that no direction within that plane is privileged. Both components are regressed against the same longitudinal $S_3$, so that the ratio of the resulting exponents is $\zeta_p^\perp/\zeta_p^\parallel$ directly (Fig.~\ref{fig:iso}). At second order the two are close, $\zeta_2^\perp/\zeta_2^\parallel\simeq0.89$, compatible with unity within the fit uncertainties, and the difference grows with order, down to $\simeq0.77$ at $p=6$. Because the low-order statistics are dominated by the most energetic, largest-scale motions, a genuine geometric anisotropy of the flow would manifest already at second order. Its absence there shows that the cascade is isotropic at the scales that dominate the low-order moments, so the growing departure at high order cannot be ascribed to a global directional bias. It reflects instead the intrinsic differential intermittency of the transverse increments, which are more intermittent than the longitudinal ones even in isotropic turbulence \citep{2002PhFl...14.1065Gotoh,2009AnRFM..41..165Ishihara}. It should in any case be treated with the caveat that the limited perpendicular extent of the $4\times2\times2~\mathrm{cMpc}/h$ region leaves the large-separation transverse pairs sparsely sampled.

\section{Longitudinal velocity structure functions}
\label{app:long_VSF}

\begin{figure}[h]
     \centering
     \includegraphics[trim=0 0 900 22,clip, width=\linewidth]{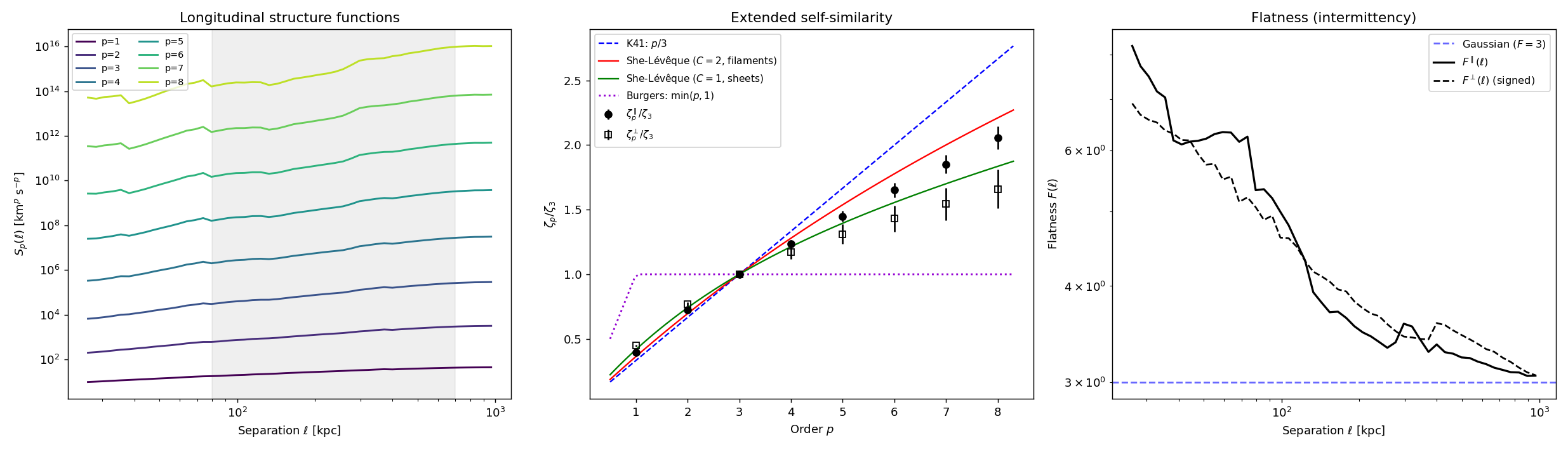}
\caption{Longitudinal velocity structure functions $S_p(\ell)$ of the diffuse
filament gas at $z=0$, for orders $p=1$ (bottom) to $p=8$ (top), computed
directly on the Voronoi cells. The shaded band marks the fit range
$[80,700]~\mathrm{ckpc}/h$. Units are $\mathrm{km}^p\,\mathrm{s}^{-p}$.}
\label{fig:vsf}
\end{figure}

Figure~\ref{fig:vsf} shows the longitudinal velocity structure functions $S_p(\ell)$ of the diffuse gas at $z=0$, from $p=1$ to $p=8$, computed directly on the Voronoi cells (Sect.~\ref{sec:vsf}). They follow clean power laws over the fit range at every order, which confirms that the diffuse gas sustains a well-defined inertial range and supports the direct fit of the third-order exponent used in Sect.~\ref{sec:cascade}.

We also verified that the scaling exponents are robust to two analysis choices. First,
because the moving mesh has near-equal cell masses, a pair count samples the
flow approximately mass-weighted. Repeating the measurement with pairs weighted
by cell volume, $w_{ij}\propto(\rho_i\rho_j)^{-1}$, so as to approximate a
volume-weighted average, changes the relative exponents $\zeta_p/\zeta_3$ by less
than $1\%$ up to fourth order. Second, restricting the extended self-similarity
fit to the range $[80,700]~\mathrm{ckpc}/h$ changes them by a comparable amount.
In both cases the shifts stay within the scatter of the orders we interpret and
grow only at the highest, poorly constrained orders, confirming that our
conclusions do not depend on the pair weighting nor on the fitting window.

\section{The enstrophy source-term budget}
\label{app:enstrophy}

\begin{figure}[h]
\centering
\includegraphics[width=\linewidth]{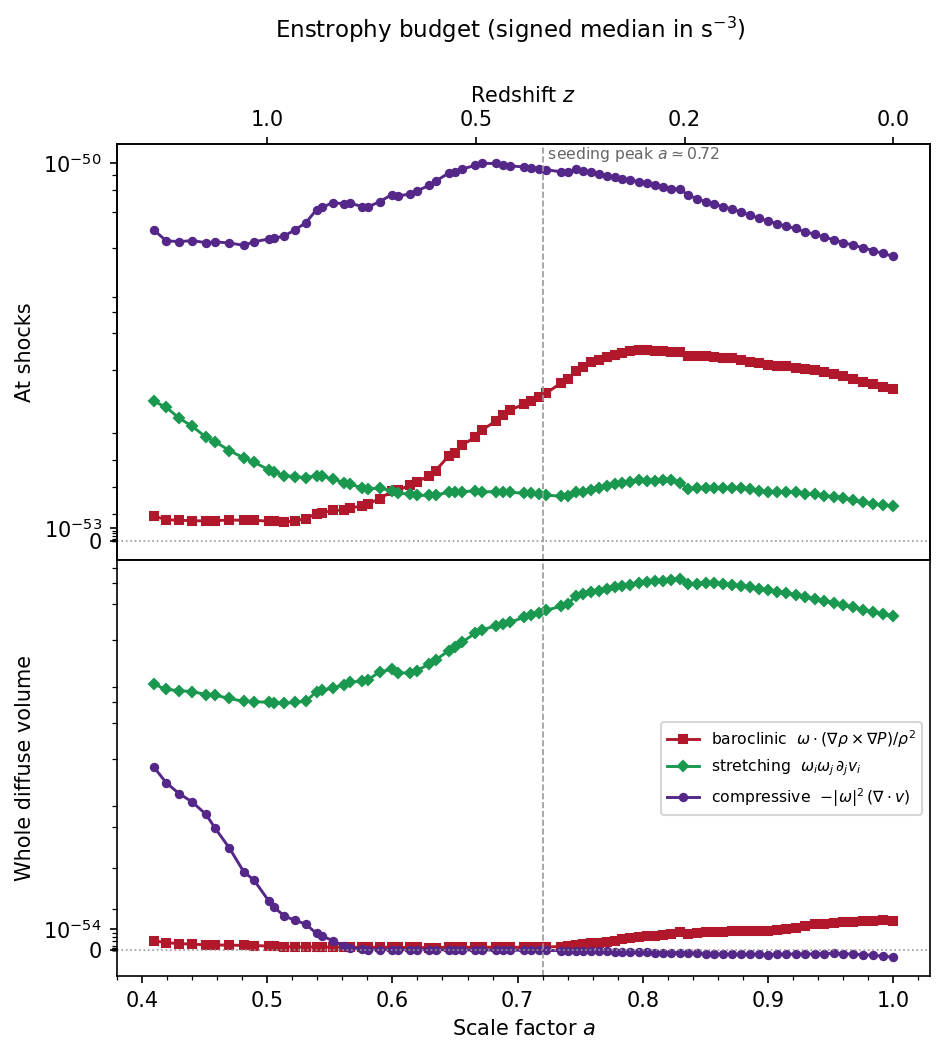}
\caption{Signed median of the three enstrophy source terms of the diffuse gas against
scale factor $a$, in CGS units. Top: at the accretion shocks, the converging cells with
$\nabla\cdot\boldsymbol{v}<0$. Bottom: over the whole diffuse volume. The dashed
line marks the $a\simeq0.72$ joint peak of the whole-volume enstrophy and
baroclinic medians (Fig.~\ref{fig:time_evo}); the shock-restricted baroclinic
source peaks slightly later, near $a\simeq0.8$.}
\label{fig:enst_budget}
\end{figure}

To move from the co-location of Sect.~\ref{sec:vort} and Sect.~\ref{sec:growth} to
a term-by-term attribution, we decompose the source of the enstrophy
$\tfrac{1}{2}|\boldsymbol{\omega}|^2$ into its three physical channels, the baroclinic
term $\boldsymbol{\omega}\cdot(\nabla\rho\times\nabla P)/\rho^2$, the vortex-stretching
term $\omega_i\omega_j\partial_j v_i$, and the compressive term
$-|\boldsymbol{\omega}|^2(\nabla\cdot\boldsymbol{v})$. Figure~\ref{fig:enst_budget} shows
the signed median of each channel against scale factor, at the accretion shocks (the
converging cells with $\nabla\cdot\boldsymbol{v}<0$) and over the whole diffuse volume.

The three channels play distinct roles. The baroclinic term is the only one that
generates vorticity from an irrotational flow, since the other two are proportional to
the vorticity already present. Its median rises through the assembly and peaks near
$a\simeq0.8$ at the shocks, tracking the epoch of oblique accretion of
Sect.~\ref{sec:vort}. The compressive term is the largest source at the shocks, where
converging gas amplifies the vorticity already there, but it does no net work over the
diffuse volume, where it averages to zero once expansion is included. The
vortex-stretching term dominates the whole diffuse volume from $a\simeq0.6$ onwards and
remains the leading source at $z=0$, which is the signature of a developed cascade whose
strain field feeds the enstrophy.

The budget thus demonstrates the sequence inferred in the main text. The baroclinic term
seeds the vorticity at the oblique accretion shocks during the assembly, the compressive
term amplifies it locally at those shocks without sustaining it, and the vortex-stretching
of the developed cascade sustains it thereafter.

\section{Continuous evolution of the intermittency}
\label{app:timeinterm}

\begin{figure}[h]
     \centering
     \includegraphics[trim=0 0 0 0,clip, width=\linewidth]{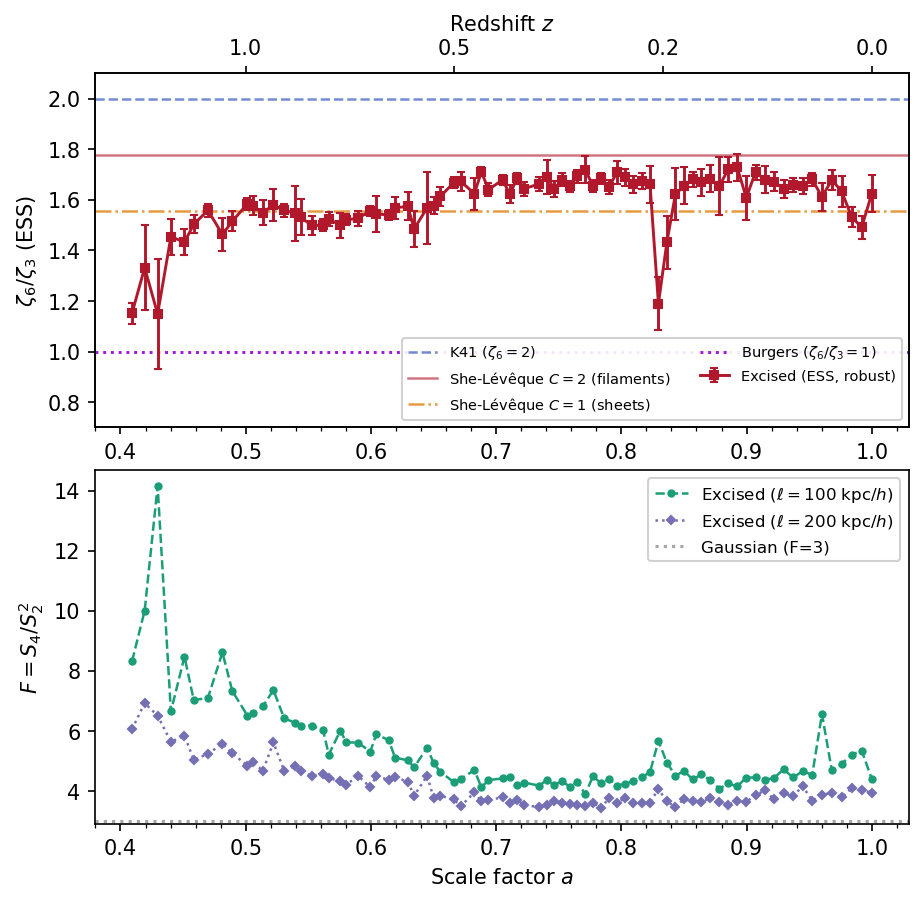}
\caption{Redshift evolution of two intermittency diagnostics of the diffuse
filament gas from $z\simeq1.4$ to $z=0$, on the same selection as the $z=0$
analysis. Top: relative exponent $\zeta_6/\zeta_3$ from a robust
extended-self-similarity fit, with the Kolmogorov, She--L\'ev\^eque ($C=1$,
$C=2$) and Burgers references. Error bars are the fit uncertainties. Bottom: scale-dependent flatness
$F=S_4/S_2^2$ at $\ell=100$ and $200~\mathrm{ckpc}/h$, with the Gaussian value
$F=3$.}
\label{fig:timeinterm}
\end{figure}

The evolution described at three epochs in Sect.~\ref{sec:cascade} is continuous
across all snapshots (Fig.~\ref{fig:timeinterm}). Both the relative sixth-order
exponent $\zeta_6/\zeta_3$ and the scale-dependent flatness at fixed separation
describe the same trend, the intermittency decreasing from a strongly
shock-dominated regime at early times towards a developed, mildly intermittent
cascade that settles closest to the sheet-like $C=1$ model by $a\simeq0.7$.

\end{appendix}

\end{document}